\documentclass[11pt,a4paper]{article}
\usepackage{amssymb} \usepackage{amsmath} \usepackage{graphicx}
\usepackage{epsfig,latexsym}
\begin{document}

\def\bef{\begin{figure}}
\def\eef{\end{figure}}
\newcommand{\ans}{ansatz }
\newcommand{\be}[1]{\begin{equation}\label{#1}}
\newcommand{\beq}{\begin{equation}}
\newcommand{\ee}{\end{equation}}
\newcommand{\beqn}[1]{\begin{eqnarray}\label{#1}}
\newcommand{\eeqn}{\end{eqnarray}}
\newcommand{\bd}{\begin{displaymath}}
\newcommand{\ed}{\end{displaymath}}
\newcommand{\mat}[4]{\left(\begin{array}{cc}{#1}&{#2}\\{#3}&{#4}
\end{array}\right)}
\newcommand{\matr}[9]{\left(\begin{array}{ccc}{#1}&{#2}&{#3}\\
{#4}&{#5}&{#6}\\{#7}&{#8}&{#9}\end{array}\right)}
\newcommand{\matrr}[6]{\left(\begin{array}{cc}{#1}&{#2}\\
{#3}&{#4}\\{#5}&{#6}\end{array}\right)}
\newcommand{\cvb}[3]{#1^{#2}_{#3}}
\def\lsim{\raise0.3ex\hbox{$\;<$\kern-0.75em\raise-1.1ex
e\hbox{$\sim\;$}}}
\def\gsim{\raise0.3ex\hbox{$\;>$\kern-0.75em\raise-1.1ex
\hbox{$\sim\;$}}}
\def\abs#1{\left| #1\right|}
\def\simlt{\mathrel{\lower2.5pt\vbox{\lineskip=0pt\baselineskip=0pt
           \hbox{$<$}\hbox{$\sim$}}}}
\def\simgt{\mathrel{\lower2.5pt\vbox{\lineskip=0pt\baselineskip=0pt
           \hbox{$>$}\hbox{$\sim$}}}}
\def\unity{{\hbox{1\kern-.8mm l}}}
\newcommand{\eps}{\varepsilon}
\def\ep{\epsilon}
\def\ga{\gamma}
\def\Ga{\Gamma}
\def\om{\omega}
\def\omp{{\omega^\prime}}
\def\Om{\Omega}
\def\la{\lambda}
\def\La{\Lambda}
\def\al{\alpha}
\newcommand{\ov}{\overline}
\renewcommand{\to}{\rightarrow}
\renewcommand{\vec}[1]{\mathbf{#1}}
\newcommand{\vect}[1]{\mbox{\boldmath$#1$}}
\def\tm{{\widetilde{m}}}
\def\mcirc{{\stackrel{o}{m}}}
\newcommand{\Dm}{\Delta m}
\newcommand{\dm}{\varepsilon}
\newcommand{\tanb}{\tan\beta}
\newcommand{\nbar}{\tilde{n}}
\newcommand\PM[1]{\begin{pmatrix}#1\end{pmatrix}}
\newcommand{\up}{\uparrow}
\newcommand{\down}{\downarrow}
\def\omE{\omega_{\rm Ter}}
%

\newcommand{\Dsusy}{{susy \hspace{-9.4pt} \slash}\;}
\newcommand{\DCP}{{CP \hspace{-7.4pt} \slash}\;}
\newcommand{\mc}{\mathcal}
\newcommand{\gr}{\mathbf}
\renewcommand{\to}{\rightarrow}
\newcommand{\gtc}{\mathfrak}
\newcommand{\wh}{\widehat}
\newcommand{\br}{\langle}
\newcommand{\kt}{\rangle}


\def\lsim{\mathrel{\mathop  {\hbox{\lower0.5ex\hbox{$\sim$}
\kern-0.8em\lower-0.7ex\hbox{$<$}}}}}
\def\gsim{\mathrel{\mathop  {\hbox{\lower0.5ex\hbox{$\sim$}
\kern-0.8em\lower-0.7ex\hbox{$>$}}}}}

\def\nn{\\  \nonumber}
\def\de{\partial}
\def\brf{{\mathbf f}}
\def\bbf{\bar{\bf f}}
\def\bF{{\bf F}}
\def\bbF{\bar{\bf F}}
\def\bA{{\mathbf A}}
\def\bB{{\mathbf B}}
\def\bG{{\mathbf G}}
\def\bI{{\mathbf I}}
\def\bM{{\mathbf M}}
\def\bY{{\mathbf Y}}
\def\bX{{\mathbf X}}
\def\bS{{\mathbf S}}
\def\bb{{\mathbf b}}
\def\bh{{\mathbf h}}
\def\bg{{\mathbf g}}
\def\bla{{\mathbf \la}}
\def\bmu{\mathbf m }
\def\by{{\mathbf y}}
\def\bmu{\mbox{\boldmath $\mu$} }
\def\bsig{\mbox{\boldmath $\sigma$} }
\def\bunity{{\mathbf 1}}
\def\cA{{\cal A}}
\def\cB{{\cal B}}
\def\cC{{\cal C}}
\def\cD{{\cal D}}
\def\cF{{\cal F}}
\def\cG{{\cal G}}
\def\cH{{\cal H}}
\def\cI{{\cal I}}
\def\cL{{\cal L}}
\def\cN{{\cal N}}
\def\cM{{\cal M}}
\def\cO{{\cal O}}
\def\cR{{\cal R}}
\def\cS{{\cal S}}
\def\cT{{\cal T}}
\def\eV{{\rm eV}}
%




\large
 \begin{center}
 {\Large \bf Neutron Majorana mass from exotic instantons} \
 \end{center}

 \vspace{0.1cm}

 \vspace{0.1cm}
 \begin{center}
{\large Andrea Addazi}
\footnote{E-mail: \,  andrea.addazi@alice.it}
\end{center}
{\it \it Dipartimento di Fisica,
 Universit\`a di L'Aquila, 67010 Coppito AQ and
LNGS, Laboratori Nazionali del Gran Sasso, 67010 Assergi AQ, Italy}

 \begin{center}
{\large Massimo Bianchi}
\footnote{E-mail: \, massimo.bianchi@roma2.infn.it}
\end{center}
{\it Dipartimento di Fisica, Universit\'a di Roma ÒTor VergataÓ
I.N.F.N. Sezione di Roma ÒTor VergataÓ, Via della Ricerca Scientifica, 1
00133 Roma, ITALY}

\vspace{1cm}
\begin{abstract}
\large
We show how a Majorana mass for the neutron could result from non-perturbative quantum gravity effects peculiar to string theory. In particular, ``exotic instantons" in  un-oriented string compactifications with  D-branes extending the (supersymmetric) standard model could indirectly produce an effective operator $\delta m \, n^t n+h.c$. In a specific model with an extra vector-like pair of `quarks', acquiring a large mass proportional to the string mass scale (exponentially suppressed by a function of the string moduli fields), $\delta m$ can turn out to be as low as  $10^{-24}-10^{-25}\, \rm eV$.

The induced neutron-antineutron oscillations could take place with a time scale $\tau_{n \bar{n}} > 10^8 s$ that could be tested by the next generation of experiments. 
On the other hand, 
proton decay and FCNC's are automatically strongly suppressed and are compatible with the current experimental limits. 

Depending on the number of brane intersections, the model may also lead to the generation of Majorana masses for R-handed neutrini. 
Our proposal could also suggest neutron-neutralino or neutron-axino oscillations, with implications in UCN, Dark Matter Direct Detection, UHECR and Neutron-Antineutron oscillations. 

This suggests to improve the limits on neutron-antineutron oscillations, as a possible test of string theory and quantum gravity. 

\end{abstract}

\baselineskip = 20pt


\section{Introduction}

Does a Majorana fermion exist in our Universe? This question remains one of the most intriguing for particle physics. When we address this issue, we would immediately think of neutrini. But curiously, Ettore Majorana suggested the {\it neutron} as a candidate rather than the neutrino \cite{1}.
A Majorana mass term $\delta m \, n^t n+h.c$  leads to neutron-antineutron oscillations
through a non-diagonal mass matrix \cite{nnbar}
\be{hnnnn}
 \mathcal{M}_{\rm eff} = \left( \begin{array}{cc} m_{n} & \delta m
\ \\ \delta m^* & m_{n} \ \\
\end{array} \right)
\ee 
with two neutron mass eigenstates $n_{\pm}=(n\pm \bar{n})/\sqrt{2}$
\footnote{We assume that CPT is not violated, the neutron and antineutron are assumed to have the same mass. However it was proposed in \cite{AJO}, that $n-\bar{n}$ could be an interesting test for CPT. The current experimental limit on the mass difference is $|m_{\bar{n}}-m_{n}|/m_{n}<10^{-5}$ \cite{Baldo}. The limit  for proton-antiproton is $|m_{\bar{p}}-m_{p}|/m_{p}<10^{-9}$ while for kaon-antikaon is $|m_{K_{+}}-m_{K_{-}}|/m_{K_{+}}<10^{-19}$.}. 
These transitions violate the Baryon number $B$,  $|\Delta B|=2$. 
So, the mystery of the existence of Majorana fermions is strictly related to another deep question: the violation of Baryon or Lepton numbers. 
The apparently ``ugly" and incomplete structure of the Standard Model of elementary particles, based on the non-semi-simple gauge group $G = SU(3)\times SU(2) \times U(1)$, displays some accidental  {\it at priori}  unexpected miracles implied by the renormalizability of its lagrangian: no Flavour Changing Neutral Currents (FCNC), no Lepton and Baryon number violations etc.  So our ``incomplete theory" automatically predicts stable baryons against proton decays, stable leptons against processes like $\mu \rightarrow \gamma e$, no neutrino-less double beta decays, no neutron-antineutron oscillations etc.
The Standard Model continues to surprise with its solidity also at the TeV-scale, directly  tested at the LHC, so one has no direct experimental indication as how, if not why, to extend it. Yet, some indirect  evidence from neutrino oscillations, dark matter and dark energy, baryo-genesis and lepto-genesis suggest the need for new physics beyond the SM. 
Neutron-antineutron oscillations could be another signal in this direction, connected not only to the question posed by Majorana, but also to B-violation and  baryo-genesis\footnote{Vafa-Witten theorem shows that the strong sector of the Standard Model cannot spontaneously break vectorial symmetries like Baryon number \cite{Vafa-Witten}. The proof is based on the exponential fall off of the fermion propagator, on the assumption that $\theta_{QCD}$ be zero and on the exclusion of any extension of the SM. In practise a theory without a mass-less Goldstone boson in the perturbative spectrum cannot produce it by binding massive particles. Lattice QCD simulations seem to support the validity of the theorem \cite{MPL1}\cite{MPL2}\cite{MPL3}. On the other hand, non-perturbative stringy instantons violate the hypothesis introducing a new scale, connected to the string theory scale, as we will see in this paper. In general no global continuous symmetries are expected to survive in a quantum theory of gravity.} 

After inflation, shaving off all hairs of the primordial Universe and restoring matter-antimatter symmetry $B=0$ and $L=0$, 
baryonic and leptonic number asymmetries could be generated by interactions, which satisfy the three Sakharov's conditions i) B-violation or L-violation, ii) CP-violation and iii) system out of thermal equilibrium \cite{Sakharov}.
These strongly motivate to believe that L and B are not really exact quantum numbers, 
but only ``accidental" symmetries of the SM, explicitly broken by non-renormalizable 
operators at the scale of the unknown new physics beyond the SM \cite{Weinberg}. 
For example, $L$-violation $(|\Delta L|=2)$ could be induced by the dimension $5$ Weinberg  term
\be{7}
\mathcal{O}^W_{\Delta L=2}=\frac{1}{\mathcal{M}}(\ell^{\alpha}\phi_{\alpha})^{t}(\ell^{\alpha}\phi_{\alpha})
\ee
with $\ell$ denoting leptons and $\phi$ the Higgs doublet. (\ref{7}) can produce Majorana masses for neutrini
$m_{\nu}\approx  \langle\phi\rangle^2/\mathcal{M}$.
For example a simple model generating (\ref{7}) is the ``see-saw mechanism", introducing a heavy RH neutrino $N$ with mass term and Yukawa $\frac{1}{2}\mathcal{M} N^2+\phi\bar{\ell}N+h.c$. Integrating out $N$ produces the scale $\mathcal{M}$ that, compatibly with the experimental limits on the neutrino masses $m_{\nu}<0.1\, \rm eV$, should be around the Grand Unification scale $\mathcal{M}\sim 10^{15 \div 16}\, \rm GeV$. 
GUT's also induce new dimension $6$ operators like $\frac{1}{\mathcal{M}_{\rm GUT}^{2}}qqq\ell$ with $\Delta (B-L)=0$, as expected for a non-anomalous global symmetry, 
allowing for proton decay via $p\rightarrow \pi^{0}e^{+}$ or $p\rightarrow K^{+}\nu$ etc.
So we could well envisage the possibility of generating dimension $9$ six-fermion operators of the type $\frac{1}{\mathcal{M}^5}(udd)^{2}$ or  $\frac{1}{\mathcal{M}^5}(qqd)^{2}$, inducing a Majorana mass for the neutron. 

\section{Neutron-antineutron oscillations}

A surprise about neutron-antineutron oscillations comes from their relatively mild experimental limits with respect to other rare processes like proton decay $\tau_{p}>10^{34}\, \rm yr$ or neutrinoless double-beta decay $\tau_{0\nu\beta\beta}>10^{25}\, \rm yr$  \cite{PDG}. Limits on $n-\bar{n}$ oscillations
are placed by experiments on beams of slow neutrons, launched along a shielded tube with a speed $v \sim 1000\,\rm m/s$ for a time interval $\Delta t \sim 0.1\,\rm s$, in a suppressed magnetic field ${\cal B}\sim 10^{-4}\,\rm Gauss$. Eventually an anti-neutron $\bar{n}$ might be detected at the end of the long tube, where its annihilations would produce typical signatures in the target. The limit on the oscillation time is $\tau_{n \bar{n}}=1/\delta m>0.86 \times 10^{8}\rm s$ with $90\%$ C.L, that implies the bound  $\delta m<7.7\times 10^{-24} \rm eV$ \cite{Baldo} on the Majorana mass. For reviews see also \cite{K}. 
This kind of experiments has an ample margin of improvement. In the near future there is the concrete possibility of increasing the neutron propagation time to $\Delta t \sim 1\,s$ and to suppress the magnetic field to ${\cal B}\sim 10^{-6}\div10^{-5}\,\rm Gauss$. Thus one could enhance the experimental limit to $\tau_{n-\bar{n}}>10^{10}\,s$ \cite{ProjectX}. 

Neutron-antineutron transitions for free neutrons at $\tau \sim 10^{8}\,\rm s$ do not lead to dangerous destabilization of nuclei.  In the atomic nucleus, one has to consider the presence of nuclear binding energies, that strongly suppress the contribution of any external magnetic fields and of the neutron or antineutron $\beta$-decay widths. The effective hamiltonian takes the form
\be{hnnnn}
 \mathcal{H}_{\rm eff} = \left( \begin{array}{cc} m_{n}-V_{n}& \delta m
\ \\ \delta m^{*} & m_{n}- V_{\bar{n}} \ \\
\end{array} \right)
\ee 
where $V_{n}$ and $V_{\bar{n}}$ are the binding energies in the nucleus for a neutron and an antineutron. $V_{\bar{n}}<<V_{n}$, $|V_{\bar{n}}-V_{n}| \sim V_{n} \sim 10 \rm MeV$. 
The neutron in the nucleus is essentially free for a time that can be estimated from the generalized uncertainty principle to be 
$$ \Delta E \Delta t \sim 1 \quad \longrightarrow \quad t_{\rm free} \sim \frac{1}{E_{\rm bind}} \sim10^{-23} \rm s $$
with $E_{\rm bind}$ the average binding energy of the nucleon in the nucleus. The oscillation probability is given by
 $$P_{n\bar{n}}=\frac{\delta m^{2}}{\delta m^{2}+ \Delta V^{2}} \sin^{2} \sqrt{\delta m^{2}+\Delta V^{2}}t \simeq \frac{4 \delta m^{2}}{(\Delta V)^{2}} \longrightarrow \tau_{A}=\frac{1}{p_{A}} \sim 10^{32} \, \rm yr$$ 
 where $\tau_{A}$ is the internuclear transition lifetime, and $p_{A}$ the transition rate. 
The limits from nuclear stability translated into free-neutron are not so different 
form the direct search ones. For Oxygen for example it is $\tau>2.4\times 10^{8}\, \rm s$ \cite{Ossigeno}, for Iron $\tau>1.3 \times 10^{8}\, \rm s$ \cite{Ferro}.

The Majorana mass is induced by effective operators of the form 
\be{Mmass}
\delta m=\langle \bar{n}|\mathcal{H}_{eff}|n\rangle =\frac{1}{\mathcal{M}}\sum_{i}c_{i}\langle\bar{n}|\mathcal{O}_{i}|n\rangle
\ee
that depend on non-perturbative strong IR dynamics.
A complete classification of the matrix elements $\langle\bar{n}|\mathcal{O}_{i} |n\rangle$ (for different Lorentz and color structures) can be found in \cite{Classification}.  Using the MIT bag model  \cite{Classification} \cite{128}, the calculations involve six-folds integrals of spherical
Bessel functions from the quark wave-functions. One can show that
\be{found}
\langle\bar{n}|\mathcal{O}_{i}|n\rangle\sim O(10^{-4})\, \rm GeV^{6}\simeq (200\, \rm MeV)^{6}\simeq \Lambda_{QCD}^{6}
\ee
More recent calculations using lattice QCD confirm these estimates \cite{134}. 
So, one can roughly estimate the Majorana mass induced by effective operators as
\be{Majorana}
\delta m \sim \left(\frac{\Lambda_{\rm QCD}}{\mathcal{M}}\right)^5 \Lambda_{\rm QCD}\sim 10^{-25}\left( \frac{1\, \rm PeV}{\mathcal{M}}\right)^5\, \rm eV
\ee
times some group theory factor, {\it viz.} Clebsch-Gordan coefficients, depending on the particular model. 
Eq. (\ref{Majorana}) tells us the limit on the new physics scale inducing the Majorana mass $\mathcal{M}>300\, \rm TeV$. In near future experiment the PeV-scale should be at reach. 

\section{Neutron Majorana mass from exotic instantons}

Henceforth, we would like to show how a Majorana mass for the neutron could indirectly
result from non-perturbative effects of quantum gravity type. 
In particular, we propose a simple un-oriented string theory model with intersecting D-branes, where 
``exotic stringy instanton effects", perfectly calculable and controllable in the case under consideration, can play this role. 
Unlike `gauge' instantons, `exotic' instantons do not admit an ADHM construction\footnote{For recent review see \cite{Bianchi:2007ft, Bianchi:2009ij, Bianchi:2012ud}}. Though subtly compatible with gauge invariance, thanks to compensating axionic shifts, they elude a natural gauge theory interpretation. In the open-string theory context all instantons, gauge or exotic, admit a simple geometric interpretation: they are nothing but special D-branes, Eucliden D-branes (E-branes) wrapping an internal cycle, that could intersect the `physical' D-branes.
In a restricted class of string compactifications with a (MS)SM-like spectrum, these effects are naturally present and explicitly computable. 
So, we would like to argue that string theory could produce observable phenomena generated by non-perturbative effects that do not exist in a gauge theory, even without large extra dimensions that would favour a TeV-scale quantum gravity. The second suggestion is that phenomenological aspects of string theory could be simpler to test in rare processes and in particular in neutron physics rather than at colliders. 

Obviously, a Majorana mass for the neutron could be generated in other ways, not directly related to string theory, in models that extend the standard model with GUT groups, Left-Right symmetric extensions, R-breaking MSSM or R-breaking NMSSM and so on. For a review of these, see \cite{M2}. For example in \cite{BM1} an $SO(10)$ GUT model without supersymmetry is suggested, that with a more  complicated multiplet structure can achieve exact unification, also increasing the  life-time of the proton to $\tau_p \sim 10^{34}\, \rm yr$. Assuming that color-sextet scalars survive down to the TeV-scale -- so much so that LHC would discover them -- diquark couplings of these scalars lead to neutron-antineutron oscillations. A similar model cannot be simply accommodated within open un-oriented string theory \footnote{On the other hand, (supersymmetric) $SO(10)$ models can be easily constructed within heterotic string theory \cite{100}-\cite{101}-\cite{102}-\cite{103}, or F-theory \cite{105}-\cite{106}-\cite{107}, but they are less appealing, less simple and less controllable.}.

Alternatively, R-parity breaking MSSM's are consistent with several string inspired models. 
But if one allows for all R-parity breaking renormalizable terms in the MSSM, like $h_{UUD} \epsilon^{ijk} U^c_iU^c_j D^c_k$,  $h_{LQD} L^\alpha Q_\alpha^i D_i^c$ and $h_{HLE} H_u^\alpha L_\alpha E^c$ (neglecting the family structure) then one needs a severe and unnatural fine tuning of the parameters to avoid  proton decay with $\tau_{p}<10^{34}\, \rm yr$ \footnote{R-parity violating operators have unnaturally small couplings but they are allowed by gauge invariance. Other R-parity violating gauge-invariant non-renormalizable effective operators that one could consider are $QQQH$ or $LHLH$.}. The proton decay constraint does not give much room for $n-\bar{n}$ oscillations at $\delta m \sim (10^{8\div 10}\, \rm s)^{-1}$. In general R-parity breaking seems to complicate rather than solve the phenomenological problems of  the MSSM. In particular,  it introduces 48 extra dangerous parameters wrt the R-parity-conserving MSSM case.
As an alternative, one can give up supersymmetry altogether and introduce a sort of ``RH-neutron" that via a see-saw mechanism could induce a Majorana mass for the neutron \footnote{These considerations are briefly summarized, in a footnote, in the paper \cite{dodi}}. This last mechanism cannot be embedded -- at least in a straightforward fashion -- in a string inspired SM-like model or in a supersymmetric GUT. 

In the same class of SM-like string inspired model as in the present investigation, but with a more direct mechanism, exotic string instantons can also generate a Majorana mass for the RH neutrino as proposed in \cite{Blu1, Ibanez1, Ibanez2}. The Majorana mass for the RH neutrino $N$ is given by $M_{N}\sim M_{S}e^{-S_E}$, where $M_{S}$ is the string scale and $S_E$ measures the (complexified) world-volume of the exotic instanton brane in string units and depends on the moduli fields.

These seem to be the only simple possibilities to generate a neutrino or a neutron mass without Left-Right symmetry or explicit R-parity violating (non-)renormalizable terms. Exotic instantons naturally lead to {\it dynamical} R-parity breaking in  MSSM, inducing R-violating {\it non-renormalizable} effective operators. 
In particular, as we will see in the next section, with a simple construction one can explain within this paradigm not only why R-parity violating operators are naturally suppressed by high-scale mass powers, but also how one can avoid a proton decay faster than $10^{34\div 35}\, \rm yr$ and $n-\bar{n}$ oscillations faster than $10^{8}-10^{10}\, \rm s$ \footnote{Another possibility for neutrino mass generation is within large extra dimension scenari \cite{ADDR}, that {\it mutatis mutandis} could work also for neutrons.}. 
Let us mention {\it en passant} that the $\mu$-term problem in the MSSM could also be solved thanks to exotic stringy instantons as proposed in \cite{Ibanez1}. 

The string models that one can consider in order to embed (N-MS)SM-like theories, with chiral matter and interesting phenomenology,  are divided in three classes: 
i) type I with magnetized D9-branes wrapping a $CY_{3}$ or alike; ii) un-oriented type IIB with  space-time filling D3-branes and D7-brane wrapping holomorphic divisors in a  $CY_{3}$; iii) un-oriented type IIA with intersecting D6-branes, wrapping 3-cycles in $CY_{3}$. 
In the last class of models, the different particle families and tri-linear couplings arise from double and triple intersections, respectively. The interactions can be derived in a direct way from string amplitudes and the low-energy limit can be naturally described by matter coupled ${\cal N}=1$ SUGRA, with chiral and vector multiplets. 
The remarkable feature that motivates our paper is the presence of non-perturbative stringy effects in the effective action. 
{\it Gauge} instantons, that are point-like configurations in the 4d Minkowski space, in (un-)oriented type IIA, correspond to Euclidean D2 (E2) branes wrapping the same 3-cycle as a stack of ``physical" D6-branes. 
The D6/D2 system has 4 mixed ND directions and the ADHM construction is obtained from open strings. 
In type I, one has E5 branes in the internal space,  with the same magnetization as the D9, that are wrapped on the entire $CY_{3}$. In (un-)oriented IIB one has D-instantons E(-1) or E3 wrapping the same holomorphic divisor as a stack of ``physical" D7-branes. 

On the other hand, {\it exotic} instantons correspond in type IIA to E2 branes, that are still point-like in the 4d Minkowski space but wrap different 3-cycles from the ``color" D6 branes. These are not ordinary gauge instanton configurations: there are no ADHM-like constraints, no bosonic moduli in the mixed sectors and the number of mixed ND directions is typically 8. 
The counterpart in type I are E5 branes wrapping the entire $CY_{3}$, but with different magnetization from the D9's, or E1 wrapping holomorphic cycles. In (un-)oriented type IIB with D3- and D7-branes they are E3 wrapping different holomorphic divisors from the D7's. 
\footnote{For an overview of instanton effects in strings theory see: \cite{uno}-\cite{sei}  for world-sheet
instantons, \cite{NS5braneHet, Bianchi:1994gi, Bianchi:1996zj} for NS5-brane and ALE instantons, \cite{sette}-\cite{nove} for E2-instantons in the Type IIA theory,
\cite{dieci}-\cite{dodici} for M2-brane and M5-brane instantons in M-theory, \cite{tredici}-\cite{quindici}  for the D3-
D(-1) system, \cite{sedici} for the effect of background fluxes on E2-instantons , \cite{diciassette} for E3-instantons in Type IIB theory, for Heterotic / Type I duality \cite{Bianchi:1998vq, Bianchi:2007rb}.}

The D-brane construction depends on whether the strings are oriented or un-oriented.
For oriented strings, a stack of $N$ D-branes, parallel to each other, supports a $U(N)$ gauge group. For example, for a theory of type IIA compactified on a six-dimensional manifold $\mathcal{M}$, a particular configuration is given by $K$ stacks of intersecting D6-branes filling the 4-dimensional Minkowski spacetime and wrapping internal `Lagrangian' 3-cycles $\Pi_{a}$ of $\mathcal{M}$. 
 The open string degrees of freedom  give rise to the gauge theory on the D6-brane world-volumes. 
 There are two sectors: states with both ends on the same stack and those connecting
different stacks of branes. 
The latter include chiral fermions
living at each four-dimensional intersection of two stacks of D6-branes $a$ and $b$ and transforming in the 
bi-fundamental representation of $U(N_{a} )\times U(N_{b})$ \cite{sessantadue}.
The number of intersections of two branes $a$ and $b$, ${\cal I}_{ab}=[\Pi_{a}]\cdot[\Pi_{b}]$ is a topological invariant giving the multiplicity of massless fermions times a sign depending on the chirality. 
On the other hand, the closed strings can propagate in the entire 10 dimensional space-time and account for gravitational fields, axions and scalar moduli fields. 

When the D-branes are space-time filling, $\Omega$-planes are introduced that are necessary for tadpole cancellation \cite{Bianchi:1990yu, Bianchi:1990tb, Bianchi:1991eu, sessantatre, sessantaquattro, MBJFM} and the consistency of the theory. $\Omega$-planes combine world-sheet parity with a (non) geometric involution in the target space. As a result Left- and Right-moving modes of the closed strings are identified. Both closed and open strings become un-oriented and more choices for the gauge groups and their representations are allowed \cite{Bianchi:1990yu, Bianchi:1990tb, Bianchi:1991eu}. D-branes come in two different types. 
There are branes whose images under the orientifold action $\Omega$
are different from the initial branes, and also branes that are their own
images under the orientifold projection. 
Stacks of the first type combine with their mirrors 
and give $U(N)$ gauge groups. Stacks of the second type give $SO(N)$
or $Sp(2N)$ gauge groups.
In this context, we could embed realistic gauge groups with chiral matter in a globally consistent model \cite{Angelantonj:1996uy, Angelantonj:1996mw}. A simple way to construct a local SM-like model with open (un-)oriented strings is to consider a simple intersecting D-brane configuration, with 4-stacks, schematically represented in Fig.~1. 
This corresponds to a SM extension as
$U(3) \times U(2) \times U(1) \times U(1)$ or alternatively $U(3)\times Sp(2)_{L}\times U(1)_{L}\times U(1)_{I_{R}}$ \cite{quarantuno}. 
In the next section we will present the basic features for the mechanism generating a Majorana mass for the neutron. Later on we will discuss relevant aspects of the model such as suppressed proton-decay or neutron-neutralino (or neutron-axino) mixings. 
For the time being, let us stress that E-branes are subject to the $\Omega$-projections very much like the `physical' D-branes. In particular we will be interested in E2-branes which are `transversely' invariant under $\Omega$ and carry an $O(1)$ gauge group. These and only these carry the minimal number of fermionic zero-modes (two) required for the generation of a dynamical super-potential rather than some higher-derivative F-term.

\section{A simple model}

\begin{figure}[t]
\centerline{ \includegraphics [height=8cm,width=0.8 \columnwidth]{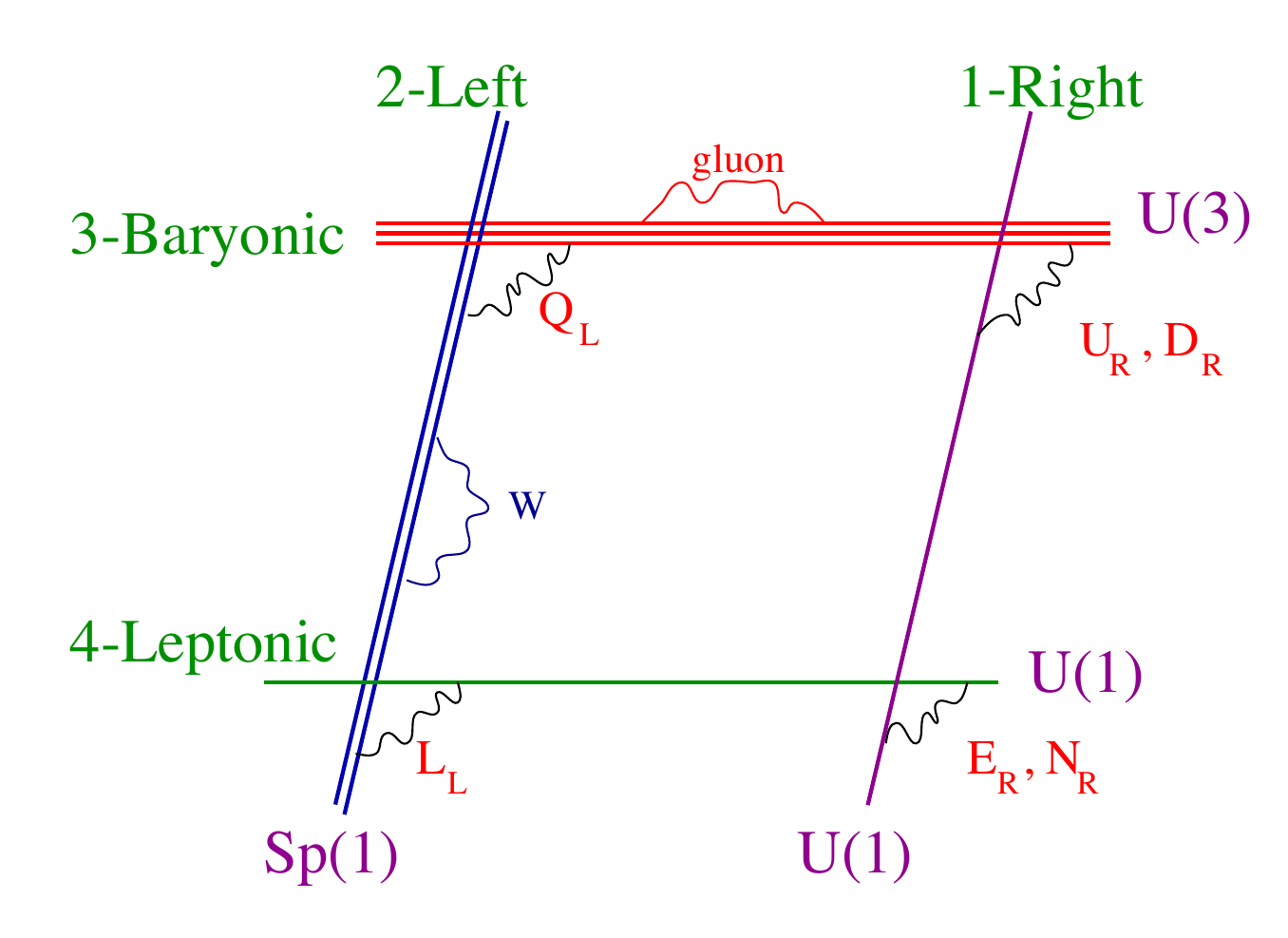}}
\vspace*{-1ex}
\caption{Schematic representation of the $U(3) \times Sp(2) \times U(1) \times U(1)$ D-brane model \cite{Anchordoqui:2012wt}. Alternatively, one could consider $U(3)_{B}\times U(2) \times U(1)_{b}\times U(1)_{c}$ (with $Sp(2) \rightarrow U(2)$).}
\label{plot}   
\end{figure}

\begin{figure}[t]
\centerline{ \includegraphics [height=6cm,width=0.8 \columnwidth]{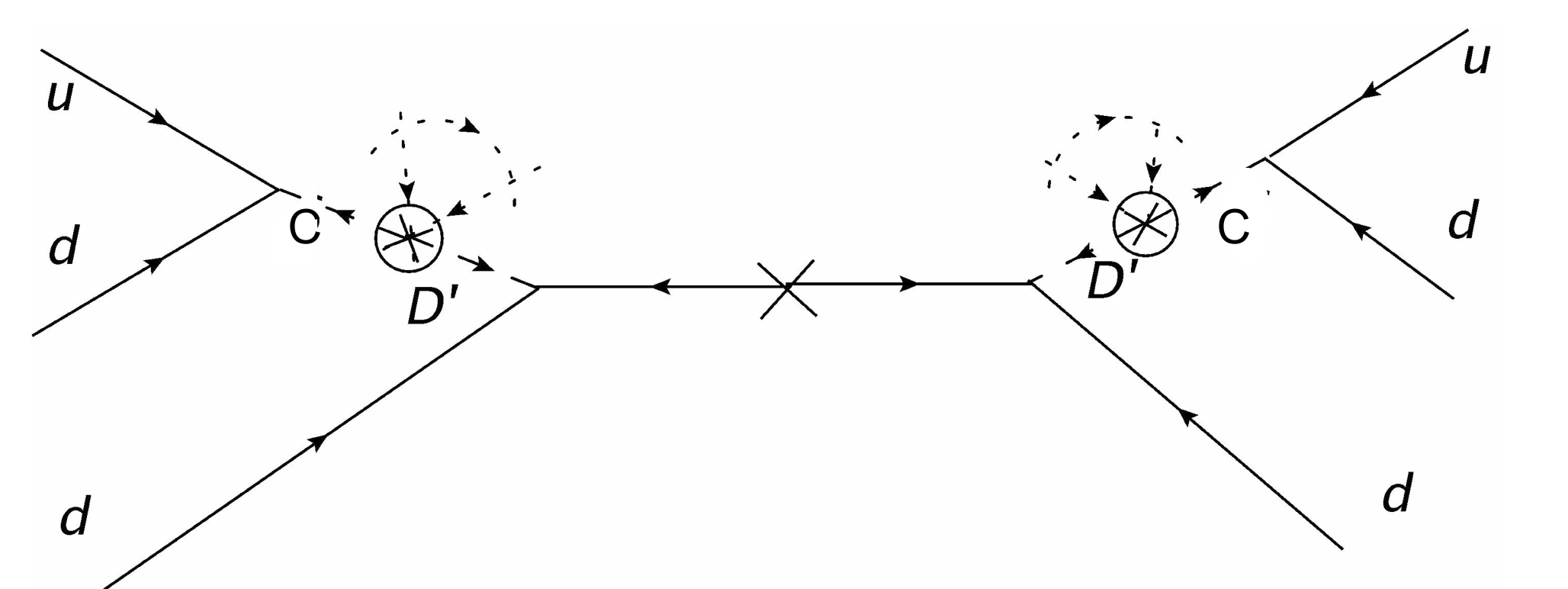}}
\vspace*{-1ex}
\caption{The diagram inducing neutron-antineutron oscillation: 
$C_{-2/3}$ and $D'_{+2/3}$ form the new vector pair, 
mixing through non-perturbative stringy instanton effects (white crosses).
The higgsino in the propagator can connect the two specular parts of the diagram 
through a Majorana mass term (in general there is an elaborate mixing between higgsini, photino, zino and wino, the mass eigenstates are called neutralini and chargini).}
\label{plot}   
\end{figure}

\begin{figure}[t]
\centerline{ \includegraphics [height=6cm,width=0.8 \columnwidth]{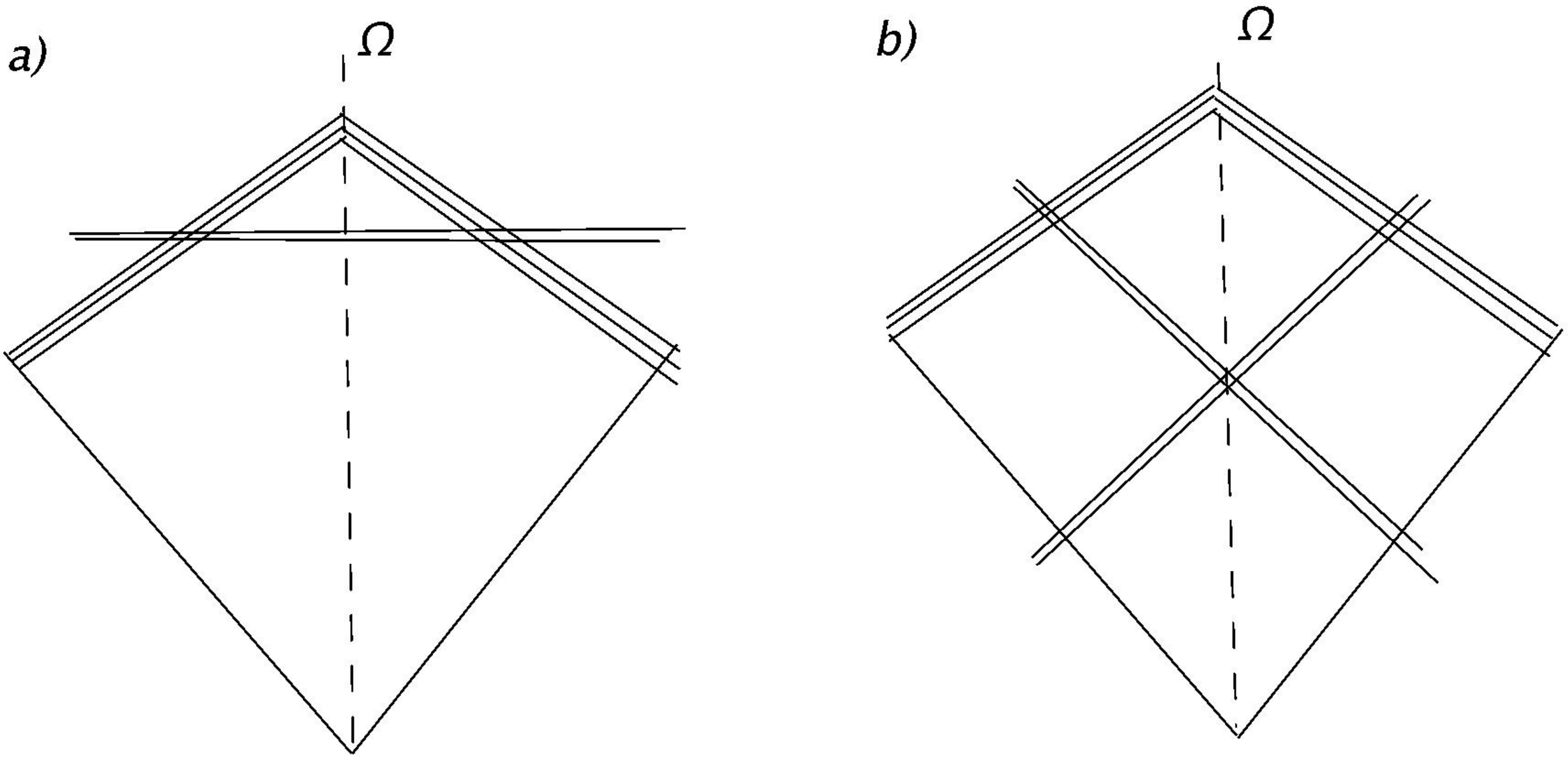}}
\vspace*{-1ex}
\caption{Simplified schemes of D-brane intersections: the stacks $U(3),U(2),U(1)$ are denoted by $3,2,1$ respectively. The main feature of this construction is the plane $\Omega^{-}$ reflecting the two $U(3)$'s into one another and generating the vector-like pair $C,D'$. In particular $C$ is an open string strectched between $U(3)$ and its image $U(3)'$, while $D'$ come from the fourth intersection between $U(3)$ and $U(1)$. As regards $U(2)$, the two possibilities are represented in a) and b): depending on whether or not $U(2)$ is invariant under $\Omega^{-}$. }
\label{plot}   
\end{figure}

\begin{figure}[t]
\centerline{ \includegraphics [height=6cm,width=0.9 \columnwidth]{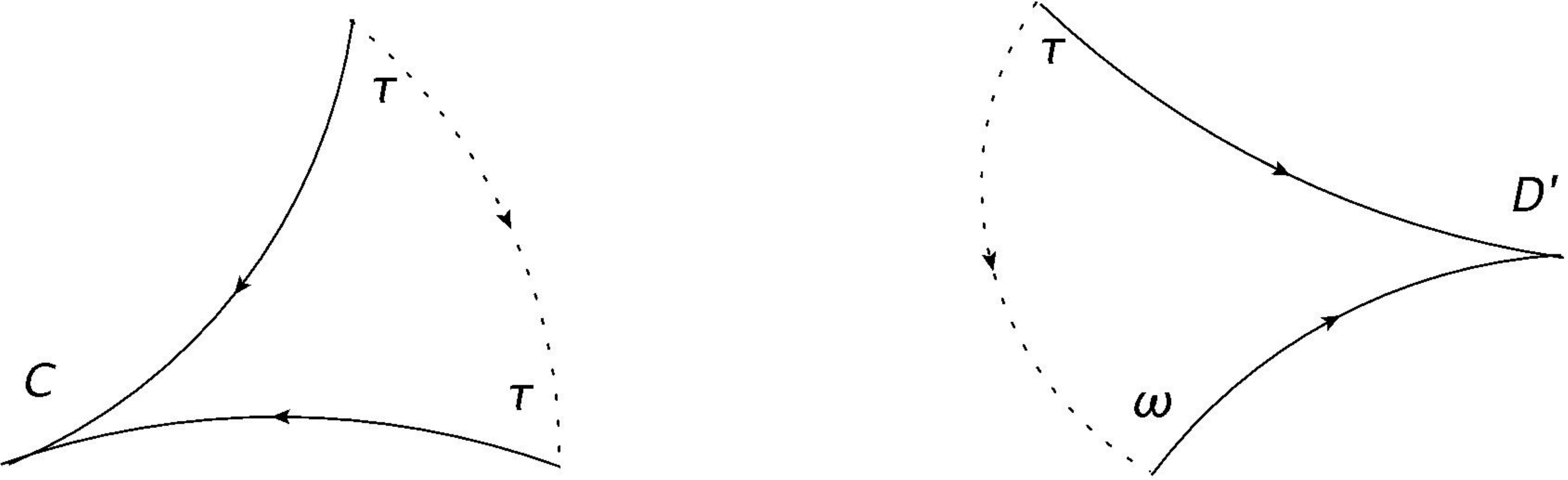}}
\vspace*{-1ex}
\caption{The two relevant mixed disk amplitudes, generating the non-perturbative coupling $\sim \epsilon_{ijk}C^{ij}D'^{k}$. $\omega,\tau$ are the four modulini interacting with $C$ and $D'$.}
\label{plot}   
\end{figure}

Let us introduce the minimal superfield content of the MSSM
\be{super}
Q_{+1/3}^{i,\alpha}, L_{-1}^{\alpha}
\ee
$$U^{c}_{i,-4/3},\,\, E^{c}_{+2},\,\, D_{i,+2/3}^{c}$$
$$H_{u,+1}^{\alpha},\,\,\,H_{d,-1}^{\alpha}$$
where $\alpha =1,2$ is for $SU(2)$, $i=1,2,3$ is for $SU(3)$ and the lower index is the $U(1)$ hyper-charge. For simplicity, the family structure is understood.

One usually considers the Baryon and Lepton number preserving renormalisable superpotential 
\be{WU}
\mathcal{W}=h_D H_{d}^{\alpha}Q_{\alpha}^{i}D^c_{i}+h_E H_{d}^{\alpha}L_{\alpha}E^c+h_U H_{u}^{\alpha}Q_{\alpha}^{i}U^c_{i}+\mu H_{u}^{\alpha}H_{\alpha d}
\ee
together with the soft susy breaking terms: scalar mass terms, Majorana mass terms for gaugini (zino, photino, gluini), trilinear $A$-terms, bilinear  $B$-terms. The superpotential $\mathcal{W}$ preserves R-parity. Models of this kind can be locally embedded in string theory with intersecting or magnetized D-branes. Building global models is more challenging.

In addition, we consider a vector-like pair that we call $D'^{c}_{i+2/3}$ and $C_{-2/3}^{i}=\frac{1}{2}\epsilon^{ijk}C_{jk}$. 
$D'$ is like a {\it 4th flavour} ($D'=D^c_{f=4}$) with exactly the same quantum numbers as the three $D^c_{f=1,2,3}$ of the MSSM. It appears when the relevant D-brane stacks have 4 rather than  3 intersections $
{\cal I}_{3,1} = \#U(3)\cdot U(1)=4$. Local tadpole cancellation \cite{Bianchi:1990yu, Bianchi:1990tb, Bianchi:1991eu, sessantatre, sessantaquattro, MBJFM} requires the presence of another U-like quark, $C$ that can appear at the self-intersection of the D-brane stack
$U(3)$ on an $\Omega^{-}$-plane, as shown in Fig.~3. Equivalently this can be described as the stack $U(3)$ intersecting its image $U(3)$' under  $\Omega^{-}$. The strings stretched between the two $U(3)$'s images transform according to the anti-symmetric combination
\be{3}
{\bf 3}_{-1/3}^{*}\times {\bf 3}_{-1/3}^{*} \vert_{A-S}\simeq {\bf 3}_{-2/3}
\ee
where ${\bf 3}_{-1/3}^{*}$ are the standard `quark' charges in the anti-fundamental representation of $U(3)$. This is a minimal extension of the 4-stacks model in Fig.~1 after including $\Omega$-planes. Although it is not our aim to construct a global string theory model with the desired properties, let us mention that several un-oriented string compactifications with intersecting or magnetised D-branes give rise to massless spectra with additional vector-like pairs such as the one we consider here \cite{CarloAug, Blumrev}. 

More precisely, one has to keep in mind that the hyper-charge group $U(1)_Y$ in this model is in general a combination of 4 $U(1)$'s in the gauge group
\be{gaugegroup}
U(3)\times U(2) \times U(1)_{c}\times U(1)_{d} \simeq SU(3)\times SU(2) \times U(1)_{3} \times U(1)_{2} \times U(1)_{c}\times U(1)_{d}
\ee
As a result $Y$ is a linear combination of 4 charges $q_{3,2,c,d}$. In fact the four $U(1)$'s are recombined into $U(1)_{Y}$, and other three $U(1)$s, one of which could be taken to be $U(1)_{B-L}$.

With these building blocks we can examine the process in Fig.~2 more closely. 
It involves a scalar color triplet with baryon charge $-2/3$ that can come from (\ref{3}). 
These cannot be s-quarks from $Q^i_{+1/3}$, but the exotic triplets $C^i_{-2/3}$, resulting from the intersection shown in Fig.~2, can do the job. 
The second ingredient that we desire for the process in Fig.~2 is a mass term for the vector-like pair. 
Due to the extra (anomalous) $U(1)$'s this is possible only through a  
non-perturbative $U(1)_Y$ preserving mass term $\mathcal{M}_{0}\epsilon^{ijk}D_{i}^{'c}C_{jk}$ \footnote{For the other standard flavours $D_{f=1,2,3}$, 
one cannot write a similar mass term: there is only one $C$ in our construction. 
So only (what we call) the 4th flavour $D'$ takes a non-perturbative mass compatibly with $U(1)_Y$. The other 3 remain massless at this level. See \cite{Bianchi:2009bg} for a similar situation.}
This could interplay with new perturbative interactions $h_{D}Q^{\alpha i}H_{\alpha}D_{i}^{'c}$ and $h_{C}Q^{i}Q^{j}C_{ij}$. 
One can integrate out the $D'_{i}, \tilde{C}^{i}$ pair and obtain at $E<<\mathcal{M}_{0}$ the effective operator
\be{effective}
\mathcal{W}_{eff}=h_{C}h_{D'}\frac{1}{\mathcal{M}_{0}}Q^{\alpha i}H_{\alpha}Q_{\beta}^{j}Q^{k\beta}\epsilon_{ijk}
\ee
the flavour structure is understood. 
At this point in order to complete the diagram in Fig.~2, we consider a higgsino propagating and connecting two operators (\ref{effective}). 

Exotic instantons can generate the desired non-perturbative mass term $\mathcal{M}_{0}\epsilon^{ijk}D_{i}^{'c}C_{jk}$, forbidden in perturbation theory by the $U(1)$ factor in $U(3)$, if they carry the correct number of fermionic zero-modes \cite{Bianchi:2007fx, Bianchi:2007wy}. 
In string theory a term with an antisymmetric tensor can only be generated in a non-perturbative way since it violates the $U(1)$ symmetry under which $\epsilon_{i_1\ldots i_N}$ carries charge $N$, {\it i.e.} 3 in our case. Even though one could replace $C_{ij}$ with $C^{k}=\epsilon^{ijk}C_{ij}/2$ and write $\mathcal{M}_{0}D'_{i}C^{i}$, $D'$ has charge $q_3 = -1$ and $C$ $q_3= -2$.

Combined with the terms (\ref{effective}), this dynamically breaks R-parity: it is not possible to identify a consistent transformations under R of $C$ and $D'$ and the other super-fields  in order to preserve  the R-parity in all the processes. This way of breaking R-symmetry is more convenient than an explicit way, since it does not generate all the possible renormalizable or non-renormalizable operators. 

As already mentioned, the relevant E2-brane should be transversely invariant under $\Omega$-projection and support an $O(1)$ gauge group. In addition to the 4 bosonic zero-modes corresponding to space-time translations it should carry two universal fermionic zero-modes, that play the role of the ${\cal N}=1$ chiral Grassamann coordinates $\theta$'s, as well as charged  fermionic zero-modes aka `modulini' living at the intersections with the physical D6-branes. The construction is shown in Fig.~4, that describes the intersections between the D6-branes that give rise to $C,D'$ and the instantonic E2 that meets our {\it desiderata, i.e.} two universal fermionic zero-modes ($O(1)$ instanton) and a single intersection each with the $U(3)$ stack (3 modulini $\tau^i$) and the $U(1)$ stack (1 modulino $\omega$). From mixed disk amplitudes, one can deduce the interactions between $C$, $D'$ and the modulini $\tau$ and $\omega$
\be{modulons}
\mathcal{L}_{E2-D6-D6'}\sim \omega D'_{i}\tau^{i}+C_{jk}\tau^{j}\tau^{k}
\ee
Integrating out the fermionic modulini one obtains the dynamical super-potential  
\be{final}
{\cal W}_{E2} = M_{S}e^{-S_{E2}}\int d^{3}\tau d\omega e^{\omega D'_{i}\tau^{i}+C_{jk}\tau^{j}\tau^{k}}=  M_{S}e^{-S_{E2}}\epsilon^{ijk}D'_{i}C_{jk}
\ee
where $\epsilon^{ijk}$ results from the integration $\int d^{3}\tau \tau^{i}\tau^{j}\tau^{k}$.
The mass scale is $\mathcal{M}_{0}\sim M_{S}e^{-S_{E2}}$, where $M_{S}$ is the string mass scale and ${S_{E2}}$ depends on the closed string moduli that parametrize the complexified size of the 3-cycle wrapped by E2. 

The superpotential term (\ref{effective}) generates the effective operator
\be{qqqtilde}
\frac{\tilde{q}\tilde{q}q}{\mathcal{M}_{0}}\frac{1}{M_{H}}\frac{\tilde{q}\tilde{q}q}{\mathcal{M}_{0}}
\ee
with $\tilde{q}$ squarks, $q$ quarks. The conversion of susy particles to SM particles brings in further suppressions. By power counting arguments, up to some adimensional ${\cal O}(1)$ factor, the 6-fermion effective operator that leads to a Majorana mass for the neutron reads 
\be{qqqqqq}
\frac{qqq}{\mathcal{M}_{0}^{2}}\frac{1}{M_{H}}\frac{qqq}{\mathcal{M}_{0}^{2}} \sim \delta{m} n^t n 
\ee
As mentioned in the introduction, the actual strength of the coupling and the value of $\delta{m}$ depend on strong IR dynamics that is beyond the scope of our analysis. Based on phenomenological models and numerical simulations \cite{Classification} \cite{128} \cite{134} one can argue that the present model can generate the effective operator $\frac{1}{\mathcal{M}^{4}}(udd)^2$ with $\mathcal{M}= (\mathcal{M}_{0}^{4}M_{\tilde{H}})^{1/5}$. The experimental bound 
$\delta m <10^{-23}\, eV$ implies $\mathcal{M}>300\, \rm TeV$.
So, one can play with $\mathcal{M}_{0}$ and $M_{\tilde{H}}$ in order to generate a value of $\mathcal{M}$ at the bound $\mathcal{M}\sim 300\, \rm TeV$. 
For instance, the choice $\mathcal{M}_{0}=M_{\tilde{H}}=300\, \rm TeV$ automatically saturates the bound. 
However higgsini (or their mixtures with wini, photini and zini in chargini and neutralini) at $100\,  \rm GeV - 10\,  \rm TeV$ scale remain a potentially interesting scenario for colliders such as LHC. In this case one needs $\mathcal{M}_{0}=700-2000\, \rm TeV$ at least. 
In both these cases, we do not need large extra dimensions and low string tension 
$M_{S}=10^3-10^4\, \rm TeV$. Since $\mathcal{M}_{0}$ is equal to the string mass times a exponentially suppressed function of the moduli, that naturally creates a hierarchy between the string mass and the $C-D'$ mass. On the other hand, a string scale of $M_{S}=10^{3}-10^{4}\, \rm TeV$ could be interesting for other rare processes and in order to alleviate the hierarchy problem of the Higgs boson. 
Finally, one can have large extra dimension in the $1-10\, \rm TeV$ range if the exponential factor is of order $1$, as discussed in \cite{ADD1} \cite{AADD} (for interesting astrophysical consequences of TeV-scale gravity see \cite{ADD2}). In this case $\mathcal{M}_{0}\sim 1-10\, \rm TeV$ and the vector like pairs would be accessible at LHC. This last possibility leads to higgsini with $M_{\tilde{H}}\sim 10^{6\div 10}\, TeV$, in contrast with susy at the TeV scale for LHC, or split-supersymmetry \cite{split} with TeV-scale quantum gravity.

\section{Further implications}

The construction we propose leads to interesting questions and implications that we cannot refrain from commenting on:

\subsection{Proton decay}
Proton decay in our model is more suppressed than in models with explicit R-parity violating terms, depicted in Fig.~5 \cite{Csaki:2013jza}. Apparently, the proton decay seems to pose a problem also in our case. The effective super-potential operator $H_{u}\epsilon_{ijk}Q^i Q^j Q^k/\mathcal{M}_{0}$ generated by exotic instantons, interplaying with the standard $H_{d} L E$, $H_{d}QD^c$ and $\mu H_{u}H_{d}$ terms, gives rise to operators like
$QQQLE/\mathcal{M}^{2}$ or $QQQQD^{c}/\mathcal{M}^{2}$.
However, a deeper analysis shows that this operators are really irrelevant for proton decay. All the $\mathcal{M}^{2}$ suppressed diagrams lead to $p$ decay channels with at least one superpartner, naturally not energetically allowed, see Fig.~7. There is no possible diagram mediated by operators of dimension $9$ competitive with $n-\bar{n}$ mixing in transition rate, only diagrams mediated by higher dimension operators exist in this model. 

\begin{figure}[t]
\centerline{ \includegraphics [height=4cm,width=1.0 \columnwidth]{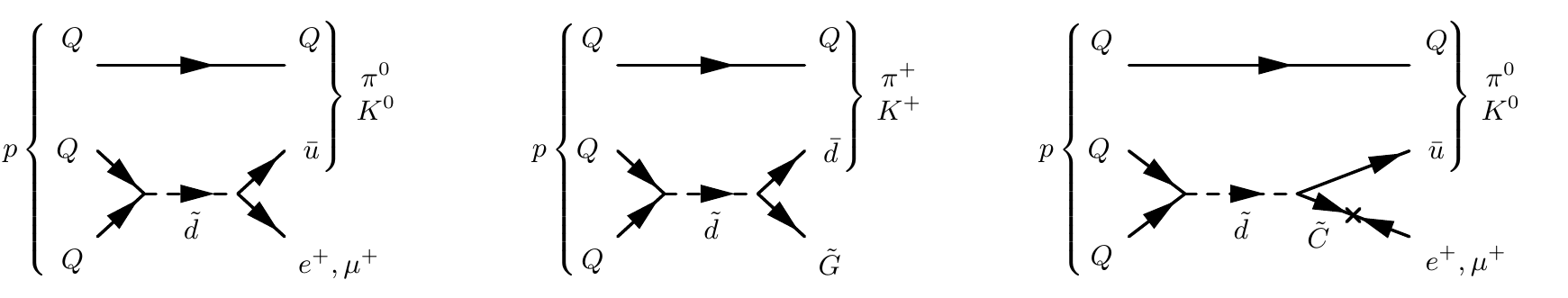}}
\vspace*{-1ex}
\caption{Proton decay in R-violating MSSM models \cite{Csaki:2013jza}. Proton decay strongly constrains the parameters of the operators involved in neutron-antineutron transitions (Fig.~6). These are automatically suppressed in our simple construction that breaks R-parity dynamically. }
\label{plot}   
\end{figure}

\begin{figure}[t]
\centerline{ \includegraphics [height=3.5cm,width=0.5 \columnwidth]{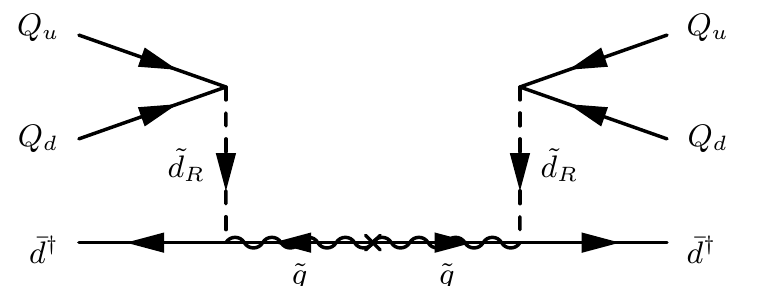}}
\vspace*{-1ex}
\caption{Neutron-Antineutron transitions from R-violating renormalizable operators \cite{Csaki:2013jza}. Our model does not generates this diagram, but the alternative one in Fig.~2. }
\label{plot}   
\end{figure}

\begin{figure}[t]
\centerline{ \includegraphics [height=5cm,width=0.7 \columnwidth]{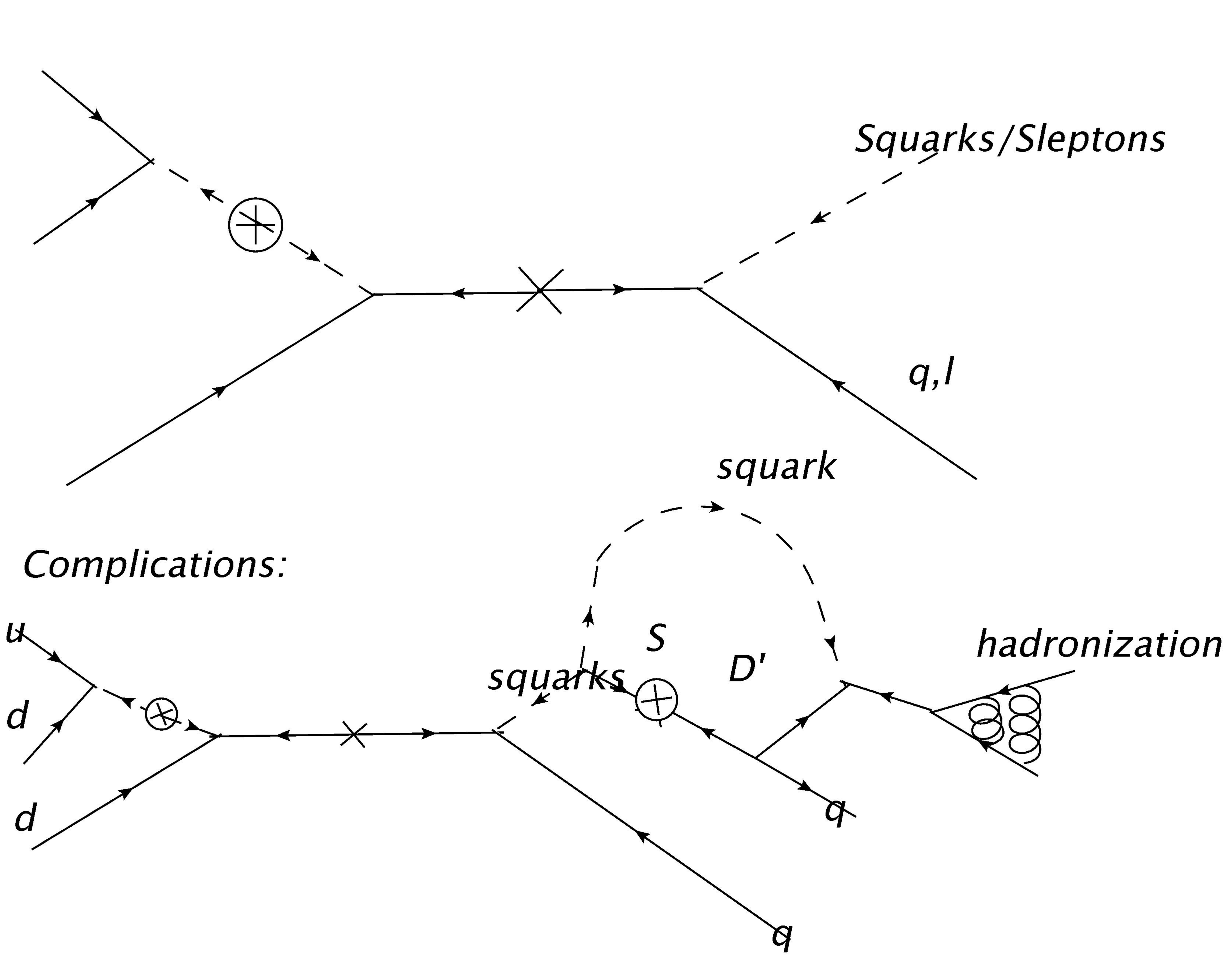}}
\vspace*{-1ex}
\caption{Diagram associated to  operators $QQQLE/\mathcal{M}^{2}$ or $QQQQD^{c}/\mathcal{M}^{2}$: these cannot induce proton decay. Implications of this diagram in higher-dimension operators can be considered. However, these are strongly suppressed with respect to $n-\bar{n}$ mixing.}
\label{plot}   
\end{figure}

\subsection{Neutralino-neutron mixing and more} 

The non-pertubatively generated effective operator $ H_{d} QQQ/\mathcal{M}_{0}$ curiously implies 
neutralino-antineutron, antineutralino-neutron, neutralino-neutron mixing (Fig.~9).
Higgsini mix with wini, photini and zini.  The resulting mass matrix has 6 mass eigenstates: 4 neutralini and 2 chargini. 
The mass terms for the neutralini read
\be{Ln}
\mathcal{L}=-\frac{1}{2}\left(\bar{\lambda}_{B_{R}},\bar{\lambda}_{3_{R}},\bar{\Psi}^{c}_{H_{1}^{0 R}},  \bar{\Psi}^{c}_{H_{2}^{0 R}}\right) \mathcal{M}_{\rm eff} \left(\bar{\lambda}_{B_{R}},\bar{\lambda}_{3_{R}},\bar{\Psi}^{c}_{H_{1}^{0 R}},  \bar{\Psi}^{c}_{H_{2}^{0 R}}\right)^T +h.c
\ee
where $\lambda_{B}$ is the gaugino associated with $B_{\mu}$ of $U(1)_{Y}$, $\lambda_{3}$ the gaugino associated to $A_{\mu}^{3}$ and $\Psi_{H_{1,2}}$ the Higgsini. The mass matrix is given by
\be{hnnnn}
 \mathcal{M}_{\rm eff} = \left( \begin{array}{cccc} M_{1} & 0 & M_{z}\cos \beta \sin \theta_{W} & -M_{Z}\sin \beta \sin \theta_{W}
\ \\ 0 & M_{2} & -M_{Z}\cos \beta \cos \theta_{W} & M_{Z}\sin \beta \cos \theta_{W}\ \\
M_{Z}\cos \beta \sin \theta_{W} & -M_{Z}\cos \beta \cos \theta_{W} & 0 &  -\mu \ \\
-M_{Z}\sin \beta \sin \theta_{W} & M_{Z}\sin \beta \cos \theta_{W} & -\mu & 0 \ \\ 
\end{array} \right)
\ee 
where $M_1$ and $M_2$ are respectively the $U(1)_Y$ and $SU(2)_L$ soft supersymmetry breaking gaugino mass terms. The eigenstates are usually denoted by $\chi_{1,2,3,4}^{0}$. In general, the mass matrix could be extended when extra $U(1)$'s appear as in our model by including axini $\tilde{a}$ \cite{SM2, CorKiri}. On the other hand, 
one has also to consider the operator $H_{d} QQQ/\mathcal{M}_{0}$, this modifies the matrix, giving rise to an effective mixing of neutrons with axini and neutralini. 
The limits on neutron oscillations in invisible channels are only of $\tau_{n-inv} > 414\, \rm s$  at $90\%\,CL$ in suppressed magnetic field \cite{UCN} \cite{UCN2} \footnote{These limits are placed in the search for a hint of Mirror Dark Matter. The phenomenology of neutron-mirror neutron oscillations are considered in \cite{oo} \cite{ot}. Currently, there is an anomaly of $5\sigma$ (with respect to the null hypothesis) in condition of magnetic field $\mathcal{B}\simeq 0.2\, \rm Gauss$ in Ultra Cold Neutron (UCN) \cite{UCN2}. This remains to be confirmed in future experiments. This could be explained if the Earth itself is the origin
of a long range Yukawa type Òfifth forceÓ acting on the neutralini or axini. In this case, the transition probability could be enhanced in condition of strong magnetic field around $0.2\, \rm Gauss$ as a resonance between the experimental magnetic field and the new interaction. }. So, there is no phenomenological problem with neutrons oscillating into the stable lightest neutralini or stable axini. Naturally, the transition probabilities will be suppressed if the neutron mass is much smaller or much larger than the neutralini and axini masses. 

On the other hand, as shown in Fig.~9, it seems that in this way the transition probabilities $\chi-n$ and $\chi-\bar{n}$ could be exactly equal, leading to a rapid transition $n-\bar{n}$ in $2-1000\, \rm s$. Clearly, if neutralini or axini have masses of $m_{\chi,\tilde{a}}>>10\, \rm GeV$, transitions into neutron and antineutron are strongly suppressed and the problem is closed, without any implication for UCN physics. However, the two transition rates could be very different if one considers the full $n\times n$ mass matrix mixing neutrons, neutralini, antineutrons, axini. In fact, in general, this matrix can violate CP, because of the Yukawa-like couplings inside the processes and extensions of the two matrix blocks $\chi-n$ and $\chi-\bar{n}$ with $N$ axini. In particular, the introduction of $N$ axini introduces  new free parameters, as non-diagonal mass terms $U_{\tilde{a}_{1,2,..,N}-n}$, CP-violating phases $\phi_{1,2,..,N}$ and axini masses $m_{\tilde{a}_{1,2,...,N}}$. Adjusting the  parameters in the model, one can get the interesting case $\tau_{\chi-n}<<\tau_{\chi-\bar{n}}$. This is not so different from the proposal of extending the mass matrix of the neutrini with one or more sterile neutrini and inducing a difference in the processes $\bar\nu_{i}\rightarrow \bar\nu_{j}$ with respect to $\nu_{i}\rightarrow \nu_{j}$. On the other hand the transition $\bar{n}-n$ through oscillations with neutralini and acini becomes an alternative to generate $n-\bar{n}$ oscillation to be tested in near future experiments.

\subsection{WIMPs and DAMA}

Light neutralini or axini are WIMP's (weakly interacting massive particles) and could be natural Dark Matter candidates or at least account for a fraction thereof.  For example, one could imagine the model dependent scenario of axino dark matter, with $\chi-n$ fast oscillations and neutralino decaying into an axino and an axion $\chi \rightarrow \tilde{a}a$. In this scenario one could assume $m_{\chi}\simeq m_{n} \simeq m_{\tilde{a}}$
($m_{\chi}-m_{\tilde{a}}\simeq m_{a}<<eV$). This situation is also very interesting for UHECR (Ultra High Energy Cosmic Rays), as we will see in the next section. 
So, our model could connect the ultra-cold neutron phenomenology with underground direct detection experiments. 

In the last 10 years or so, significant progress has been made in efforts to directly detect dark matter.
The DAMA/NaI \cite{DAMA} and DAMA/Libra \cite{DAMA2} experiments have obtained exciting results (see also \cite{DAMA3}). In particular, these experiments have observed an annual modulation at $9.3\, \sigma\, \rm C.L.$ \cite{DAMA4}, as expected for a signal from Dark particles. 
Different anomalies in other direct detection projects, CoGeNT \cite{CoGeNT}, CRESST-II \cite{CRESST} and recently in CDMS-II (CDMS-Si) \cite{CDMS}, seem to favor DAMA results. 
Interestingly, DAMA signal suggests light neutralino candidate in a region of masses $1-50\, \rm GeV$ \cite{DAMAneutralino} \footnote{In this analysis DAMA collaboration includes detector uncertainties in 
 quenching factors, channeling, nuclei Form Factor, Dark Matter Form Factor, Migdal effect (see for this last \cite{Migdal}) and so on. They also consider astrophysical uncertainties in the local rotational velocity  and local dark halo density near the Sun, and possible departures from the isothermal sphere model in density profiles, anisotropies of the velocity dispersion tensor and rotation of the galactic halo. Finally the possible contributions of
non-thermalized Dark Matter components to the galactic halo, such as the SagDEG
stream, or other kinds of streams as those arising from caustic halo models, are discussed in \cite{Sagitarius} \cite{caustic}. These
could change the local DM speed distribution and the local density.}. So light a neutralino is not ruled out at all by LEP, Tevatron and LHC data, the situation is strongly model dependent. The first data from LHC tend to disfavour a TeV-scale MSSM model \cite{LHC} and the desired $m_{\chi^{0}}\sim1\, \rm GeV$ for interesting oscillations is in tension with respect to the neutralino mass lower bound by the Cold DM relic abundance  $\Omega_{\chi}h^{2}\simeq (\Omega_{CDM}h^{2})$, derived in \cite{dicia}-\cite{dicio}: $m_{\chi^{0}}>7-8\, \rm GeV$.

In contrast to neutralino, the axino mass 
is unconstrained experimentally. Moreover
from the theoretical point of view, one can easily imagine it in
the few GeV range \cite{CKN}. Constraints on a light axino are 
not so rigidly related to the SUSY scale. Depending on the model, SUSY could be broken at higher scale compatibly with a light axino. 
For axino,  the parameter space is constrained by axion couplings with gluons, photons and fermions (see \cite{axionPDG} for a review about axion constraints), but neutron-axino oscillations are not directly related to axion PQ-like scale. So a light axino seems to be favored as a WIMP candidate of $1 \rm GeV$ with respect to neutralino. DAMA collaboration analysis for the neutralino \cite{DAMAneutralino} applies directly to the axino.  

\subsection{UHECR and GZK effect}

Other implications of neutron oscillations with a sterile partner like a neutralino or an axino could come for Ultra High Energy Cosmic Rays (UHECR) phenomenology. 
A possible effect of $n-\tilde{\chi}^{0}$ or $n-\chi^{0}$ or $n-\tilde{a}$ oscillations 
on the Greisen-Zatsepin-Kuzmin (GZK) cutoff
\footnote{UHE nucleons interact with the CMB
radiation field \cite{A9} \cite{A10}, there are two signatures that can be related to these: lepton pair-production  $p+\gamma_{CMB}\rightarrow e^{+}e^{-}p$ \cite{A11} \cite{A12}, and pion photo-production $p\gamma_{CMB}\rightarrow \pi^{0}p, \pi^{+}n$ 
called Greisen-Zatsepin-Kuzmin (GZK) cutoff \cite{A13} \cite{A14}. So, the position of GZK cutoff is approximately defined  by the energy where lepton pair-production and the pion photo-production rates.
The energy losses become practically equal at $E_{GZK}\simeq 50\, \rm EeV$ \cite{A15}. }
shape in UHECR could be detected \footnote{This effect is similar to the neutron-mirror neutron oscillations discussed in \cite{ZG}. However there is an important difference: the mirror neutrons in the the mirror sector $\beta$-decay into mirror protons. Then in the Mirror scenario we have also to consider the interactions of the mirror protons with mirror CMB.  From BBN limits, Mirror CMB temperature must be less then the ordinary CMB one. On the other hand, in our case neutralini or axini have not other relevant interactions with matter to consider, if they are assumed as WIMP-like particles. So the resulting effect on the GZK shape could be very different.}. In fact proton can collide with CMB photons, producing protons and $\pi^{0}$, or neutron and $\pi^{+}$, with practically the same probability $\mathcal{P}_{pp,pn}\simeq 1/2$ and a mean free path $l_{mfp}\sim 5\, \rm Mpc$. Then the produced neutrons could oscillate in a time interval $\tau \sim 1-500\, \rm s$ into neutralini or/and axini, which can propagate in the CMB without interactions. An example of an interesting, but model dependent, scenario may be as the following. Consider the case of a neutralino with $m_{\chi}\simeq m_{n}$ and an axino with a mass smaller then $m_{\chi}$ and assume that neutron-neutralino transition rate is much faster than neutron-axino one, this last much faster then axino-neutralino transition. This corresponds just to an effective mass matrix with a non-diagonal mixing terms constrained by the hierarchy: $\mu_{\chi-n}>>\mu_{\tilde{a}-n}>>\mu_{\chi-\tilde{a}}$.
Then one can have a decay of $\chi$ into one axion and one axino through the coupling photino-axino-axion. On the other hand, one can assume $\chi$ to be stable against other decays unrelated with this interaction. Assuming the rate for $\chi \rightarrow a \tilde{a}$ to be much slower than for${n\rightarrow \chi}$, such as $\tau_{\chi \rightarrow a\tilde{a}}>1000\, \rm s$, one could imagine a chain of processes as the one represented in Fig.~9 that would involve: i) $p\gamma_{CMB}\rightarrow n\pi^{+}$, ii) $n-\chi$ oscillations in $1-500\, \rm s$ iii)
$\chi \rightarrow a\tilde{a}$ in more than $1000\, \rm s$; iv) $\tilde{a}\rightarrow n$ after a length of $l_{mpr}>>1 \rm Mpc$ (also considering the very high Lorentz factor); v) neutron $\beta$-decays into protons. 
This chain leads to a very efficient propagation of protons and to a modification of the spectrum above the GZK cut-off. We would like to stress that this particular model is also connected with UCN and Dark Matter Underground Direct Detection experiments.


The total effect could be a modification of the spectrum at energy above the GZK. In \cite{chi} Auger's data, the GZK seems to 
appear shifted below in energy wrt theoretical expectations, if all UHECR were protons. 
Unfortunately Auger data have large error bars in the last 3 points from about $10^{19.9}$ to $10^{20.4}\, \rm eV$ and do not allow one to conclude whether the end-point is displaced or not wrt standard theoretical expectations. Moreover there are systematic uncertainties over the energy scale of $14\%$ ($\pm 0.06$ over $Log_{10}(E)$) that practically make it impossible to determine with precision how the energy spectrum ends.

On the other hand, looking at the Telescope Array (TA) data \cite{TAC} \cite{TAC2}
the experimental GZK cutoff seems to be above theoretical expectations,
apparently in contradiction with Auger data.
However, Auger and Telescope Array spectra are consistent within the 
systematic uncertainties (see \cite{A1} for analysis in common between the collaborations). 

Another unclear situation comes from the determination of the nuclei fractions, which are controversial and affected by a lot of uncertainties. 
Auger atmospheric depth data $\langle X_{Max}\rangle [g/cm^{2}]$, an indicator of the UHECR chemical composition, seem to suggest that the larger part of higher energy points are nuclei: 
protons seem to be suppressed at energy around $10^{19}\,\rm eV$, smaller than the GZK cutoff energy scale, also considering the large uncertainties of the energy scale mentioned above. In particular, Auger Collaboration 
claims the presence of nuclei in UHECR, with a gradual transition from light to heavy composition between $10^{18}\, \rm eV$ and $5\times 10^{19}\, \rm eV$ \cite{NewAuger}. 
If these results were confirmed, the $n-\chi^{0}$ and/or $n-\tilde{a}$ and/or $\chi-\tilde{a}$ oscillations would not affect the GZK cutoff shape. 
But these estimates are very 
model-dependent since it is necessary to extrapolate models of  
hadronic interactions to energies much higher than those at which they were tuned,  {\it i.e.} the TeVscale (LHC).
On the contrary, HiRes  \cite{HRFC} \cite{HC}  and TA \cite{TAC} \cite{TAC2}
show that the chemical composition is dominated by protons from $10^{18}\, \rm eV$ to $10^{20}\,\rm eV$. But they use a different data analysis and they have
 much less statistics with respect to Auger, therefore it is still not known if the disagreement is real or not (see \cite{A1}).

The observations that only $30\%$ of UHECR are within cones of few degrees from some known astrophysical source, like AGN, Blazars, Supernovae etc, seems to disadvantage 
the hypothesis that only protons compose UHECR at $E>10^{19}\, \rm eV$ due to the basic fact that for a proton of this energy the trajectory cannot be curved more than few degrees by an average intergalactic magnetic field. A nucleus with atomic number $Z$ is $Z$ times easier to accelerate and its trajectory to be curved with a magnetic field (see \cite{Blasi}). 
However, the propagation of UHE protons with $E>10^{19}eV$ could be more and more efficient because of neutron-neutralino and/or neutron-axino oscillations, they could come from unknown sources at cosmological distances (depending on model considered) not contained in the visible horizon. 
In this last case the angular correlation analysis could not be conclusive. 

The mechanism proposed is independent from the proton sources, which could be distant Blazars, or exotic new physics processes like
superheavy particle decays (for a review se \cite{KT}), monopole-antimonopole annihilations, cosmic strings or other topological defects (for a review see \cite{TD}), scalaron oscillations in $f(R)$ modifications of gravity \cite{Dolgov1}, and so on.

Naturally, a hybrid scenario can explain UHECR with $E>10^{19}eV$: a fraction could be 
 UHE nuclei coming from AGNs or other astrophysical known sources, and a part could be protons coming from unknown sources. 

So it would seem that the status of UHECR is still completely open. 
In future, with more statistics, error bars on the individual points 
will shrink a bit. Room for some improvement 
will come from better measurements of 
air fluorescence and so on \footnote{We are very grateful to Armando Di Matteo for  interesting comments on the experimental data about UHECR. }.
Then the observatory project 
JEM-EUSO will be sent on the International Space Station, 
with the opportunity to 
collect much more statistics, alas with poorer resolution \cite{Opportunity}.

For the moment, it seems more reliable to test neutron exotic oscillations in UCN experiments or in neutron base-lines. In particular the oscillations $n-\bar{n}-\chi_{0}$ or with axini could be studied in future neutron-antineutron experiments. 

On the other hand the limits on proton-charginos oscillations are more stringent (the limits are the same as for proton decay), but this is not necessarily connected with $n-\bar{n}$ or $\chi_{0}-n$ diagrams in the parameter space under consideration, including MSSM parameters, extra $U(1)$'s, $\mathcal{M}_{0}$ etc.

\begin{figure}[t]
\centerline{ \includegraphics [height=9cm,width=0.5 \columnwidth]{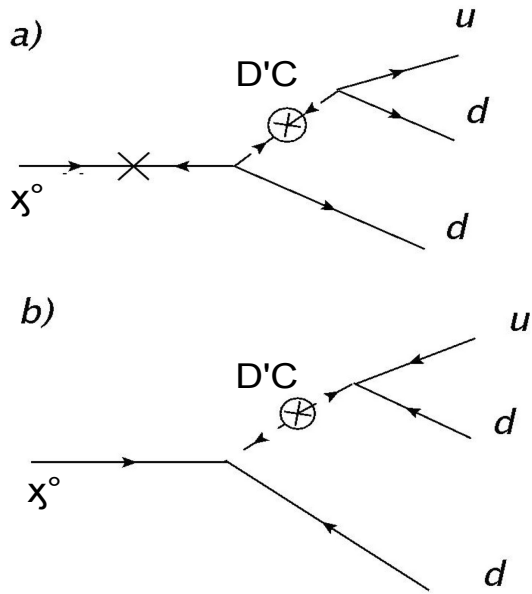}}
\vspace*{-1ex}
\caption{Diagram inducing the a) neutralinos-antineutrons (antineutralinos-neutrons) and b) neutralinos-neutrons mixings. }
\label{plot}   
\end{figure}

\begin{figure}[t]
\centerline{ \includegraphics [height=5cm,width=1.0 \columnwidth]{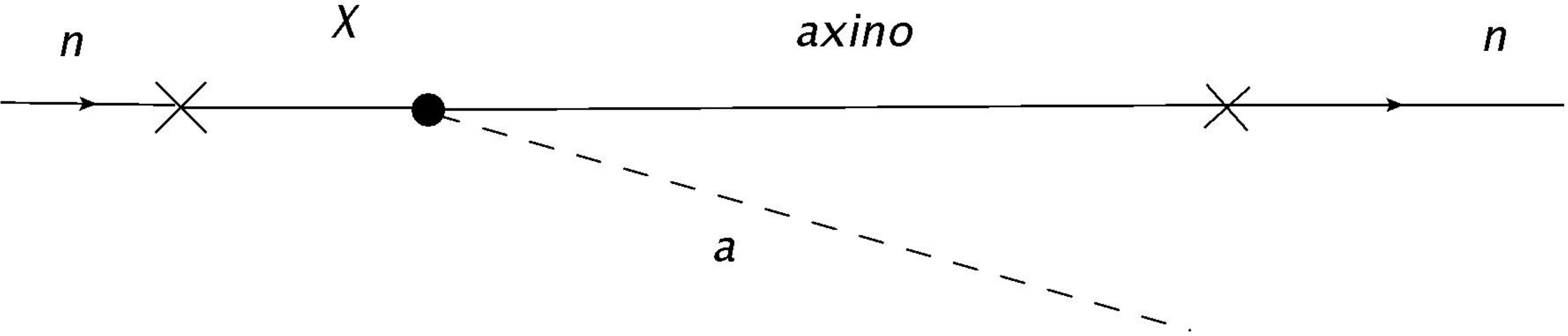}}
\vspace*{-1ex}
\caption{Example of a mechanism for UHECR protons propagation, involving rapid oscillations between neutron and neutralino $\tau_{n-\chi}\simeq 1-500\,\rm s$; neutralino decay into axion and axino with $\tau_{\chi\rightarrow a\tilde{a}}>(5\div 10)\tau_{n-\chi}$ and finally the transition of the axino into the neutron with $\tau_{\tilde{a}-n}>>\tau_{n-\chi}$.}
\label{plot}   
\end{figure}

\begin{figure}[t]
\centerline{ \includegraphics [height=9cm,width=0.7 \columnwidth]{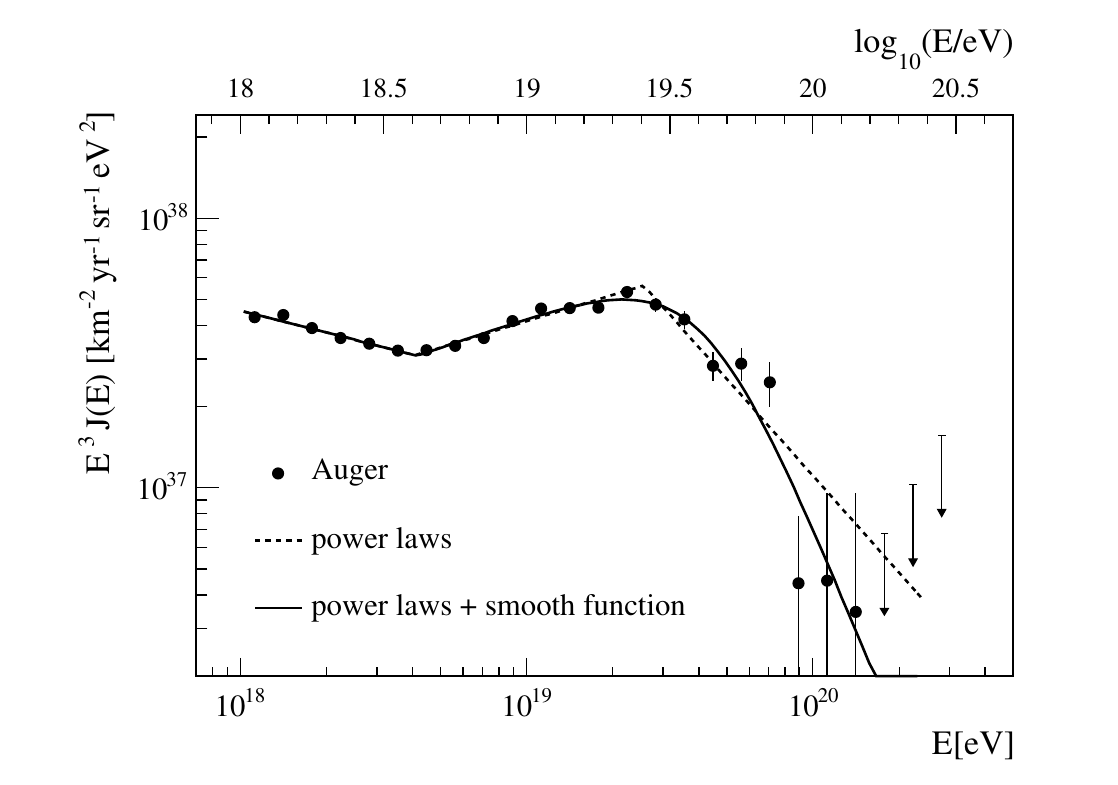}}
\vspace*{-1ex}
\caption{Pierre Auger spectrum of UHECR. In figure is also showed the best fit with three power-law models and a smooth curve. Neutron-neutralino or/and neutron-axino could change the shape of the GZK cutoff suggested by standard physics fit. In particular the end-point could be displaced at lower energies. The large error bars in the last three points could not permit to detect this effect.}
\label{plot}   
\end{figure}

\subsection{Meson physics and FCNC's} 

A natural question for phenomenology is if our model is predictive for meson physics in 
$K$, $D$, $B$, $B_{s}$ decay channels or in $K^{0}-\bar{K}^{0}$, 
 $B^{0}-\bar{B}^{0}$, $B_{s}^{0}-\bar{B}_{s}^{0}$, $D^{0}-\bar{D}^{0}$. The answer is positive, the present model can generate these processes, but they are strongly suppressed, as shown in Fig.~10 and Fig.~11.  
 \begin{figure}[t]
\centerline{ \includegraphics [height=8cm,width=0.8 \columnwidth]{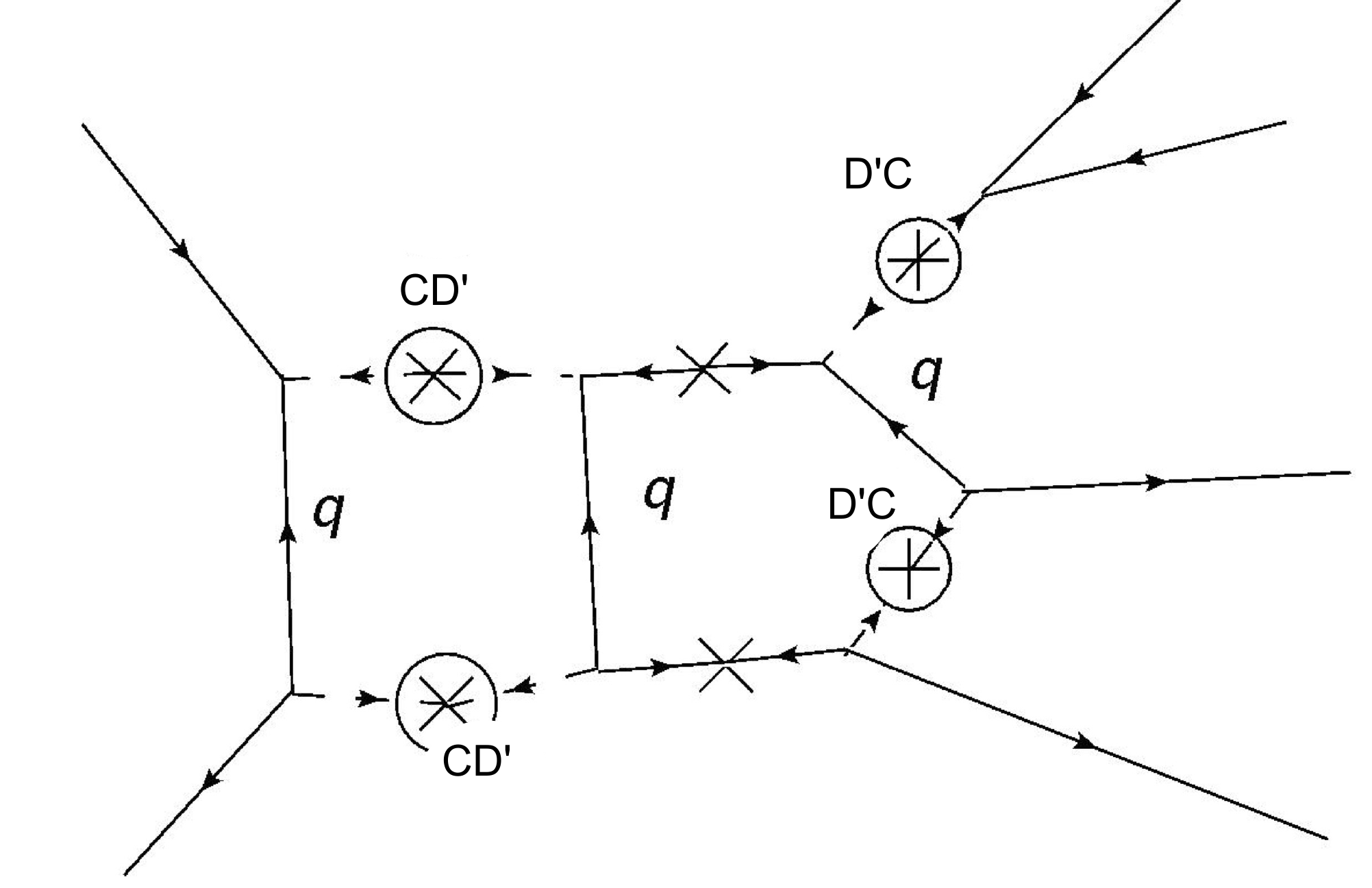}}
\vspace*{-1ex}
\caption{2-loops  diagram for meson decays in two mesons. This is mediated by two higgsinos and four $D'-C$.}  
\label{plot}   
\end{figure}
\begin{figure}[t]
\centerline{ \includegraphics [height=8cm,width=0.8 \columnwidth]{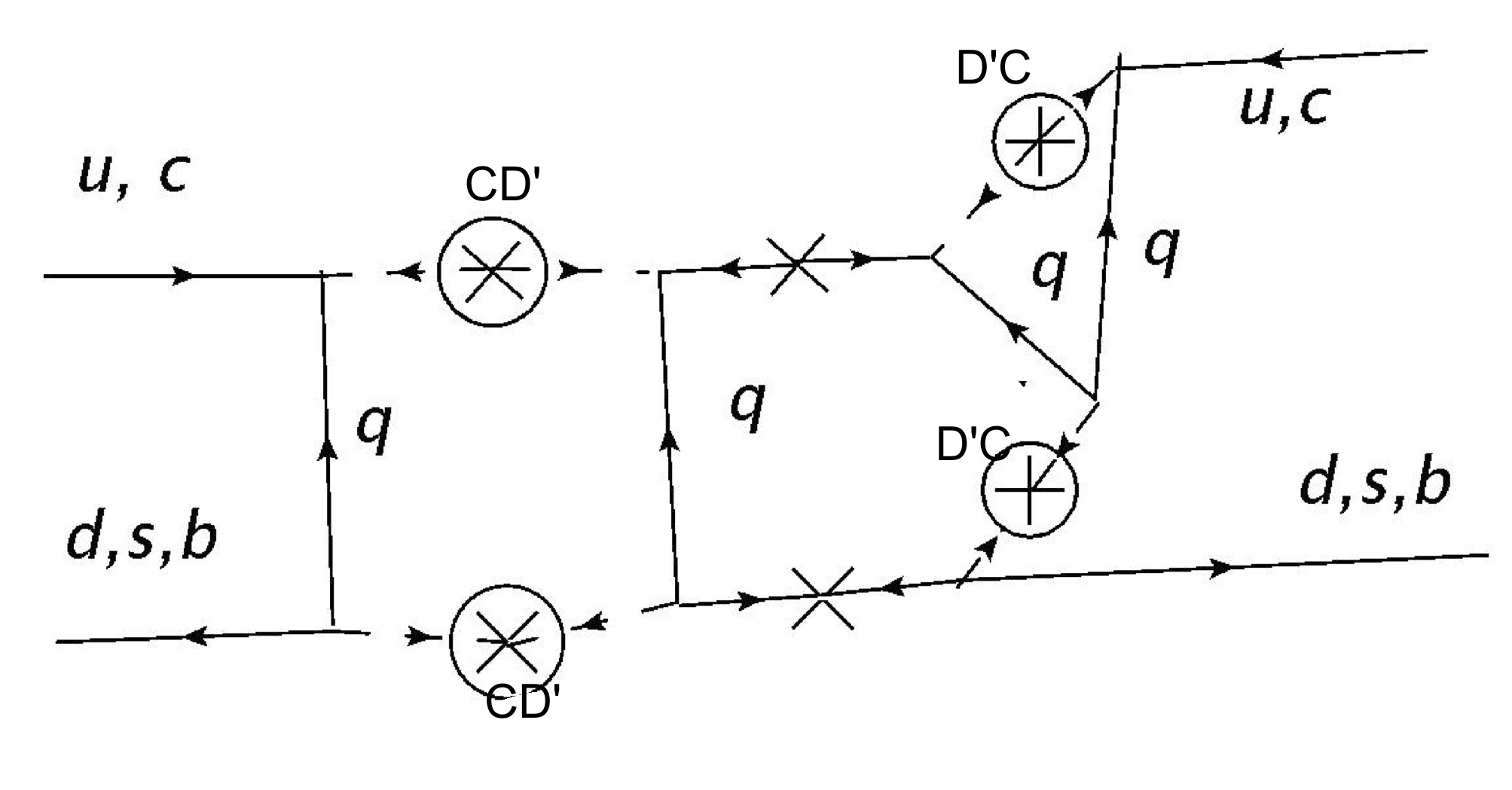}}
\vspace*{-1ex}
\caption{3-loops diagram for neutral meson-antimeson oscillation. }
\label{plot}   
\end{figure}

Another delicate question that we cannot by-pass is about FCNCs in quark sectors: are they generated in our simple model? The answer is again positive, but they are highly suppressed. Essentially, the relevant diagrams come from the variant in Fig.~11, closing one more quark-antiquark line. The 4-loop suppression is beyond any observable effects.

\subsection{Running coupling}

The introduction of $C$ and $D'$ affects the running of the strong coupling for $E>\mathcal{M}_{0}$. This does not creates a serious problem. The un-oriented open string paradigm does not seem to request gauge coupling unification at a high scale. In some sense it is a different kind of unification, including also gravity. GUT, with Supersymmetry, seems an elegant idea that 
naturally extends the Standard Model, unifying the three forces at $M_{GUT}\sim 10^{16}\, \rm GeV$. It also seems to explain why the charge is quantized, why neutrini have a small mass  and what generates fermion hierarchies and mixing angles. However GUT creates other problems like proton decay faster than the experimental limits, doublet-triplet problems for Higgs fields in the fundamental representation ${\bf 5}={\bf 3}+{\bf 2}$ and  ${\bf 5}^*={\bf 3}^*+{\bf 2}$ (giving mass of order $M_{GUT}$ to the color triplet Higgs ${\bf 3},{\bf 3}^*$ and mass $\mu$ to the Higgs doublets ${\bf 2}$) \footnote{The doublet-triplet problem can be solved in several ways. The most elegant are the missing partner or vev mechanism for SU(5) \cite{novantaquattro} and the pseudo-Nambu-Goldstone boson mechanism for
SU(6) \cite{novantacinque}. These have been shown to be consistent with gauge coupling unification and proton
decay. There are also mechanisms for explaining why the $\mu$ term is of order the SUSY breaking scale \cite{novantasei}. For some suggested
solutions in SUSY GUTs and string theory for the $\mu$ problem, see ref. \cite{novantasette} \cite{dieci} \cite{novantotto} \cite{cinquanta}. Finally, in string theory (and orbifold GUTs),  orbifold projection eliminate certain states. It has been shown
that it is possible to retain the Higgs doublets removing the Higgs triplets in
this process \cite{stringGUT}.}.
On the other hand, string theory also suggests other ways to solve these fundamental problems, by reformulating them in terms of strings, D-brane intersections, exotic Euclidean instantons and Calabi-Yau compactifications. 

\subsection{More on exotic instanton effects}  

Looking at Fig.~3, one can ask what are the consequences of ${\cal I}_{2,2'} = \#\,SU(2)\cdot SU(2)=1$ intersections. In fact these generate singlets, in analogy with the triplet $C$ from $SU(3)$, $SU'(3)$ intersections. 
In particular the construction b) proposed in Fig.~3 suggests that  twin superfields $\mathcal{L},\mathcal{L}'$ could exist. They correspond to open strings stretched between the stacks $U(2)$ and $U(2)'$. 
These could be interesting for lepton number violating processes or for flavour changing neutral currents. 

$(\mathcal{L},\mathcal{L}')$ could play also an important role in lepto-genesis. 
Finally, if their mass were around $1-10\, \rm TeV$, they could be detectable at LHC, for example in decay-channels of the singlets. 

\subsection{Different $\Omega$'s and fluxes}

Instead of an $\Omega^{-}$-plane one can consider an $\Omega^{+}$-plane in Fig.~3. This construction generates color sextets that could induce $n-\bar{n}$ oscillation in a different way, similar to the scalar color sextet of Babu-Mohapatra $SO(10)$ model \cite{BM1}. They could play an interesting role in the baryo-genesis \footnote{Sextets are also generated in a construction with $U(3)\times Sp(2)_{L}\times U(1)_{L}\times U(1)_{I_{R}}$, reflected with $\Omega^{+}$}. 
In this case the $(\mathcal{L},\mathcal{L}')$ are not antisymmetric singlets but  symmetric triplets. 

The soft susy breaking terms could also be induced by bulk fluxes. For example, 
gaugino masses could be generated by bulk fluxes such as NS-NS $H_{ijk}$ or R-R $F_{ijk}$ 3-form fluxes, from an interaction $\lambda^{t}\Gamma^{ijk}\langle\tau H_{ijk} + i F_{ijk}\rangle\lambda \sim M_{\lambda} \lambda^{t}\lambda$. So in more complicated situations, one has to consider the back-reaction of the fluxes on the ``exotic" instantons \cite{Bianchi:2011qh, Bianchi:2012kt}. These could modify the simple analysis proposed in this paper \footnote{An interesting question is whether bulk fluxes could generate -- alternatively to exotic instantons -- Majorana masses for neutrini and neutrons. Probably this is not possible. Bulk fluxes do not break any gauge invariance while gauge and exotic instantons do break $U(1)$ symmetries. Gauge instantons break anomalous (axial) $U(1)$'s, exotic instantons can break also non-anomalous (vector) symmetries like $B$.}. 

\subsection{Majorana mass for RH neutrini}

In our model, a Majorana mass terms for the RH neutrini $N_{i}$ can be generated that induces the observed small neutrino masses thanks to the see-saw mechanism. Majorana mass terms for RH neutrini are forbidden
in perturbation theory by $U(1)$ symmetries such as $U(1)_{B-L}$. However they can be generated by non-perturbative stringy instanton effects. In unoriented type IIA string models, the pseudo-scalars needed to make the $U(1)$Õs massive correspond to the R-R 3-form integrated over 3-cycles.

As we have already seen, Majorana mass terms naturally come from the intersections between an E2-instanton wrapping a 3-cycle and the background $D6$ branes wrapping different 3-cycles, see Fig. 4. 
One can derive the conditions under which an operator like $e^{-S_{E2}}NN$ can be generated \cite{Ibanez1}. This has charge $2$ under $U(1)_{B-L}$  symmetry, and charge 0 under $U(1)_{Y}$. The transformations under the (anomalous) $U(1)$ gauge symmetries could be canceled by a compensating transformations of the exponential  $e^{-S_{E2}}$, whose imaginary part is an axion with 
St\"uckelberg coupling to the $U(1)$'s. 
This conditions are compatible with our extension based on the gauge group 
$SU(3)\times SU(2) \times U(1)_{3}\times U(1)_{2}\times U(1)_{c}\times U(1)_{d}$, whereby RH neutrini come naturally from the intersections of the $U(1)$ stacks $c$ and $d$.

\subsection{Extra (anomalous) $U(1)$'s and $Z$'}

In any string-inspired extension of the (MS)SM of Fig.~1, new vector bosons $Z'$ \footnote{For discussions about the existence of
additional massive neutral gauge bosons see \cite{un}-\cite{nov}} appear that get a mass 
by a St\"uckelberg mechanism \cite{diec}. In addition, Generalized Chern-Simon (GCS) terms are introduced in order to cancel  anomalies \cite{trentacinqu}, in combination with a generalised Green-Schwarz mechanism \cite{undic}.  
If one assumes the string mass scale to be at $ M_ {S} = 1-10 \, \rm TeV $, even in our model, processes such as $Z'\rightarrow ZZ$ or $Z'\rightarrow Z\gamma$ could produce interesting signals at the LHC, as already discussed in the literature \cite{SM1}-\cite{SM2}.

\section{Conclusions and Remarks}
 
We have shown how exotic instantons can indirectly generate a Majorana mass for the neutron. The crucial ingredients are a local intersecting D6-brane configuration with $\Omega$6-planes giving rise to the MSSM super-fields plus a vector-like pair of `quark' super-fields $D', C$. An $O(1)$ instanton (E2-brane) singly intersecting  the relevant D6-branes generates a dynamical super potential mass term for $D', C$. Integrating these out, while taking into account their interactions with the standard MSSM super-fields, produces an effective Baryon number violating term that in turn leads to the desired highly-suppressed Majorana mass for the neutron. 

We have then discussed phenomenological implications and commented on potential drawbackks of the proposed mechanism. Proton decay and FCNC are highly suppressed while several signals of neutron-antineutron or neutron-neutralino/axino oscillations can give rise to interesting signatures in DM, UCN and UHECR experiments. 
This shows how interesting string theory could be for near future experiments, with its peculiar non-perturbative stringy instantons effects, not admitting a natural gauge theory interpretation. 
In particular, these could generate Majorana masses for neutrini and for neutrons.
As a consequence, the next generation of experiments on neutrinoless-double-beta decays and neutron-antineutron oscillations could test quantum gravity non-perturbative effects. In particular, limits on $n-\bar{n}$ oscillations are quite mild with respect to limits on proton decay: $\tau_{n-\bar{n}}>10^{8}\, \rm s \sim 10^{-33}\tau_{p\rightarrow \pi e, K \nu, etc.}$. 
The stringy instantons effects are completely calculable in some string models containing the Standard Model, as the one we have considered in the present paper. 
In more complicated string models as heterotic strings or in the presence of fluxes, stringy instantons effects becomes more difficult to calculate, but their existence is a quite general feature.  

We have also seen how these effects could interplay with large extra-dimensions, with a rich phenomenology for LHC. However, large extra dimensions are not necessary to generate interesting rare processes like $n-\bar{n}$ oscillations with non-perturbative stringy instantons effects. We would like to stress that  our mechanism can be compatible with highly suppressed proton decay. This is a crucial feature: if future experiments on proton decay would enhance the limits, the most interesting models for neutron-antineutron phenomenology would become models of the present kind that naturally avoid too fast a proton decay, contrary to L-R symmetric or R-violating (renormalizable) extensions of the MSSM. In fact, for these last two classes of models, an improvement on proton decay limits (for example at $10^{35}-10^{37}\, \rm yr$) 
would strongly constrain $n-\bar{n}$ oscillation at ${\cal M} \approx 300-1000\, \rm TeV$. 

We conclude that string theory could be experimentally testable in some of its Standard Model like versions, as a consequence of its better known non-perturbative aspects. Further theoretical discovery about non-perturbative aspects of string theory could show up as absolutely unique and interesting for experimental physics, in unexplored ways that one cannot imagine at present. Future experiments on rare processes as $n-\bar{n}$ could help us to clarify our understanding of the Universe and disclose its hidden Beauty.

\vspace{1cm} 

{\large \bf Acknowledgments} 
\vspace{3mm}

A..~A is thankful to Yuri Kamyshkov for the invitation at the conference "Baryon Lepton Violations" BLV 2011 (that strongly motivates him to think about the neutron-antineutron physics). We would like to thank Zurab Berezhiani, Rita Bernabei, Marco Bill\`o, Armando Di Matteo, Francesco Fucito, Mariapaola Lombardo, Askhat Gazizov, Cumrun Vafa for interesting discussions. The work of MB is partially supported by the ERC Advanced Grant n. 226455 ``Superfields" and initiated at QMUL while MB was holding a Leverhulme Visiting Professorship. 


\end{document}